\documentclass[fleqn,usenatbib]{mnras}

\usepackage{newtxtext,newtxmath}

\usepackage[T1]{fontenc}
\usepackage{ae,aecompl}
\usepackage{graphicx}	
\usepackage{amsmath}	
\usepackage{amssymb}	
\usepackage{natbib,longtable,times,multicol}
\usepackage{caption}
\usepackage[caption = false]{subfig}
\usepackage{verbatim}
\usepackage{makecell}
\usepackage{siunitx}
\usepackage{listings}

\title[Relativistic disks]{Relativistic accretion disk reflection in AGN X-ray spectra at z=0.5--4: a study of four \textit{Chandra} deep fields}

\author[]{
 \parbox[t]{18cm}{L. Baronchelli$^1$\thanks{E-mail: blinda@mpe.mpg.de}, K. Nandra$^1$, J. Buchner$^{1,2,3}$}\\\\
$^{1}$Max-Planck-Institut f\"{u}r extraterrestrische Physik, Giessenbachstrasse 1, 85748 Garching bei M\"{u}nchen, Germany\\
$^{2}$Instituto de Astrofisica, Facultad de Fisica, Pontificia Universidad Catolica de Chile, Casilla 306, Santiago 22, Chile\\
$^{3}$Excellence Cluster Universe, Boltzmannstr. 2, D-85748, Garching, Germany
}

\date{Accepted XXX. Received YYY; in original form ZZZ}

\pubyear{YYYY}

\begin{document}
\label{firstpage}
\pagerange{\pageref{firstpage}--\pageref{lastpage}}

\maketitle

\begin{abstract}
We confirm that the spectra are best fit by a model containing two Compton reflection components, one from distant material, and the other displaying relativistic broadening, most likely from the inner accretion disk. The degree of relativistic broadening indicates a preference for high black hole spin, but the reflection is weaker than that expected for a flat disk illuminated by a point source. 
We investigate the Compton reflection signatures as a function of luminosity, redshift and obscuration, confirming an X-ray Baldwin effect for both the narrow and broad components of the iron line. 
Anti-correlations are also seen with redshift and obscuring column density, but are difficult to disentangle from the Baldwin effect. 
Our methodology is able to extract information from multiple spectra with low signal-to-noise ratio, and can be applied to future data sets such as eROSITA. We show using simulations, however, that it is necessary to apply an appropriate signal-to-noise ratio cut to the samples to ensure the spectra add useful information.  
\end{abstract}




\section{Introduction}
Active galactic nuclei (AGN) are powered by matter falling onto a supermassive black hole (SMBH) \citep[e.g.][]{Rees1984}, 
via an accretion disk \citep[e.g.][]{Shakura1973,Malkan1983,Laor1989}. 

In addition to the direct disk emission AGN emit up to 20$\%$ of their bolometric luminosity in the X-ray waveband \citep[e.g.][]{Elvis1994}. The X-ray emission is produced via Compton upscattering by a hot corona \citep[e.g.][]{Haardt1991} which illuminates the innermost regions of the accretion disk. The X-ray emission can be backscattered and induce fluorescence in the inner disk \citep[e.g.][]{Fabian1989,George1991}, which leaves imprints due to the large velocities and gravitational field. Hence, the analysis of the X-ray spectrum from AGN allows us to probe general relativity and investigate the behaviour of matter in extreme gravitational fields, only a few gravitational radii from SMBHs \citep{Reynolds2003,Psaltis2008}.

In many of these systems a strong iron (Fe) K$\alpha$ fluorescent line is observed \citep{Nandra1994}. The Fe K$\alpha$ feature peaks around 6.4 keV in the rest frame. Part of the emission comes from material at scales of several parsec, most likely the torus envisaged in orientation-dependent unification schemes \citep[e.g.][]{Krolik1994,Ghisellini1994} and hence the line is relatively narrow, with velocities of a few $\sim 100$ km s$^{-1}$ \citep{Yaqoob2004,Nandra2006}.
As discussed above, the remainder of the iron K$\alpha$ feature is emitted in a region of the accretion disk in the proximity of the SMBH, and it is broadened and skewed by relativistic effects, e.g. gravitational redshift and relativistic Doppler shifts \citep{Fabian1989,Laor1991,Fabian2000,Risaliti2004}.
The shape and width of the line and the amount of broadening can help us gain information on the geometry of the system and the spin of the SMBH \citep{Brenneman2006}. Constraining the distribution of black hole spins in the whole AGN population would help determine the nature of the accretion history of the SMBH, for example by allowing us to distinguish whether the SMBH grew mostly through mergers, continuous, or "chaotic" accretion \citep{King2006,Volonteri2013}.

Strong, relativistically broadened Fe K$\alpha$ lines are observed in a number of individual, bright Seyfert galaxies in the nearby Universe, e.g. MCG-6-30-15 \citep{Tanaka1995,Fabian2002} and NGC 3516 \citep{Nandra1999}. Indeed such emission is found to be common in the nearby AGN population \citep{Nandra1997}, though the evidence for such a component is not universal \citep{Nandra2007,delaCalle2010}. 
This is probably due to fact that the strength of 
the broad reflection component is weaker than that expected from a standard accretion disc illuminated by a point source \citep{Nandra2007}, meaning that very high signal-to-noise ratio (S/N) is needed to detect it convincingly \citet{Mantovani2016,delaCalle2010}.

The fact that the reflection is weaker than expected, and differs in strength from source to source, can nevertheless reveal important information about the system such as the geometry. 
A well known effect of this type is the X-ray Baldwin or Iwasawa-Taniguchi effect \citet{Iwasawa1993} where the equivalent width of the iron line decreases with luminosity. This effect has been observed for the narrow \citep{Nandra1997b,Page2003} 
and broad \citep{Nandra1997b} components of the line. 

Clearly it is then important to characterize the iron line emission of a representative sample of AGN, including more typical objects at higher redshift. There are a few observations of relativistic broadened Fe K$\alpha$ lines at high-redshift, sometimes thanks to studies of lensed Quasars \citep[e.g.][]{Chartas2012,Dai2019}. The ubiquity or otherwise of the broad features in samples of typical AGN beyond the local Universe is hard to establish, however, given the faintness of the targets. The deepest X-ray surveys offer the opportunity to investigate this issue, and several attempts have been made to use deep surveys to infer the properties of a population of AGN and to verify the ubiquity of feature like the broadened Fe K$\alpha$ line in a population of AGN  \citep{Comastri2004,Brusa2005,Streblyanska2005}.

Most past studies have relied on stacking of large samples of low count X-ray spectra \citep[e.g.][]{Streblyanska2005,Chaudhary2012}. By stacking, one may be able to infer the population properties in cases where fitting individual spectra would not yield meaningful results. Evidence for broadening of the Fe K$\alpha$ line has been reported in several stacking studies, such as  \citet{Chaudhary2012} and \citet{Falocco2013,Falocco2014}. \citet{Corral2008}, however, combined the rest-frame spectra of 600 \textit{XMM-Newton} type-1 AGN without finding compelling evidence for a relativistic broadening of the Fe K$\alpha$ line.
Some of this work showed, however, that stacking the spectra might induce artificial broadening of the Fe K$\alpha$ in samples with a wide redshift distribution \citep{Chaudhary2012}. This problem was addressed by \citet{Falocco2012,Falocco2013,Falocco2014} and \citet{Liu2016} who compared the stacked spectra to simulations. While this approach can increase confidence in the existence of the broadened features, the simulations require an assumption about the true underlying spectrum. 

In our previous work \citep{Baronchelli2018} we employed an alternative technique to characterise the and the iron line and reflection properties of a large sample of AGN, whose X-ray spectra individually have low S/N ratio.
Instead of fitting a stacked spectrum, we used the Bayesian X-ray Analysis (BXA) software \citep{Buchner2014} to fit the individual spectra of low S/N sources. We then combined the Bayesian evidence for the putative broad reflection component to establish whether or not it was present in the whole sample. 
The analysis was performed on a sample of 199 hard X-ray selected sources from the \textit{Chandra} Deep Field South 4Ms exposure, and revealed strong evidence for a relativistically broadened X-ray reflection component from the accretion disk. The properties of the reflection also implied a preference for a maximally spinning SMBH, as compared to a non-rotating Schwarzschild black hole.

In this work, we expand the study of \citet{Baronchelli2018} to a total of four \textit{Chandra} fields.
Our aim is to investigate further the prevalence of the Fe K$\alpha$ and reflection features within typical AGN up to $z=4$, and characterize their properties. With our expanded sample, we aim to confirm our previous results, and investigate the dependence of the reflection strength with other parameters such as the luminosity, redshift and obscuration. 

The paper is structured as follows: in Sect. \ref{samplemethod}, we describe the data used in this work, and our methods of spectral and statistical analysis. The results of the work are reported in Sect. \ref{results} and interpreted and discussed in Sect. \ref{discussion}.  Sect. \ref{sumconc} summarises our results and presents our primary conclusions. 

Throughout this work, we adopt $\Omega_m$= 0.272, $\Omega_\Lambda$= 0.728, and $H_0$ = 70.4 km s$^{-1}$ Mpc$^{-1}$ \citep{Komatsu2011}.

\section{Sample and Method}
\label{samplemethod}

\begin{table*}
\caption{Number of selected sources and summed counts in in the 1--8 keV observed frame for the individual fields, and for the combined sample. The information is also given for the samples restricted to S/N$>7$ described in the text.  The source and total counts are calculated with the \textsc{Sherpa} tool \texttt{calc$\_$data$\_$sum}. }
\begin{tabular}{llccccc}
\hline \hline
& & Total & CDFS 7Ms & CDFN & AEGIS & COSMOS \\

 \hline
 \textbf{ALL}  &  &  & &  &  &    \\ 
 \hline
& Number of sources & 2237 & 199 & 376 & 540 & 1122 \\ 
& Source counts & 655951 & 313914 & 127563 & 119462 & 95011 \\ 
& Total counts & 759356 & 364635 & 148096 & 140087 & 106538 \\ 
 \hline
 \textbf{S/N $\geq$ 7}  &  &  & &  &  &    \\ 
 \hline
& Number of sources & 2165 & 198 & 349 & 539 & 1079 \\ 
& Source counts & 654899 & 313864 & 126922 & 119448 & 94663\\ 
& Total counts & 755815 & 364472 & 145604 & 140033 & 105706 \\ 
 \hline
 \hline
\end{tabular}
\label{SampleDetails}
\flushleft 
\end{table*}

\begin{figure*}
\centering
\includegraphics[width=0.9\columnwidth,trim={0cm 0.5cm 2.2cm 2.7cm},clip]{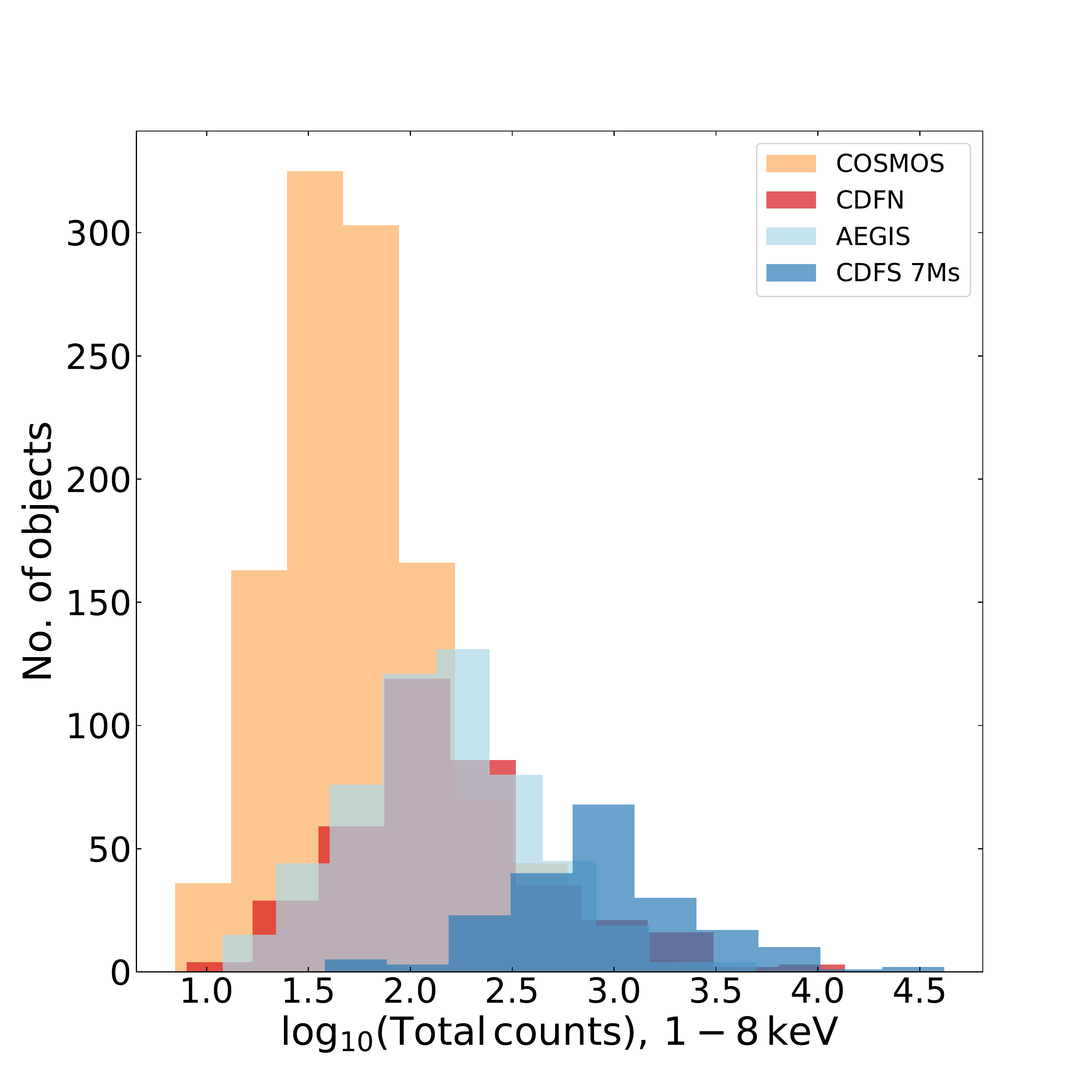}
\includegraphics[width=0.9\columnwidth,trim={0cm 0.5cm 2.2cm 2.7cm},clip]{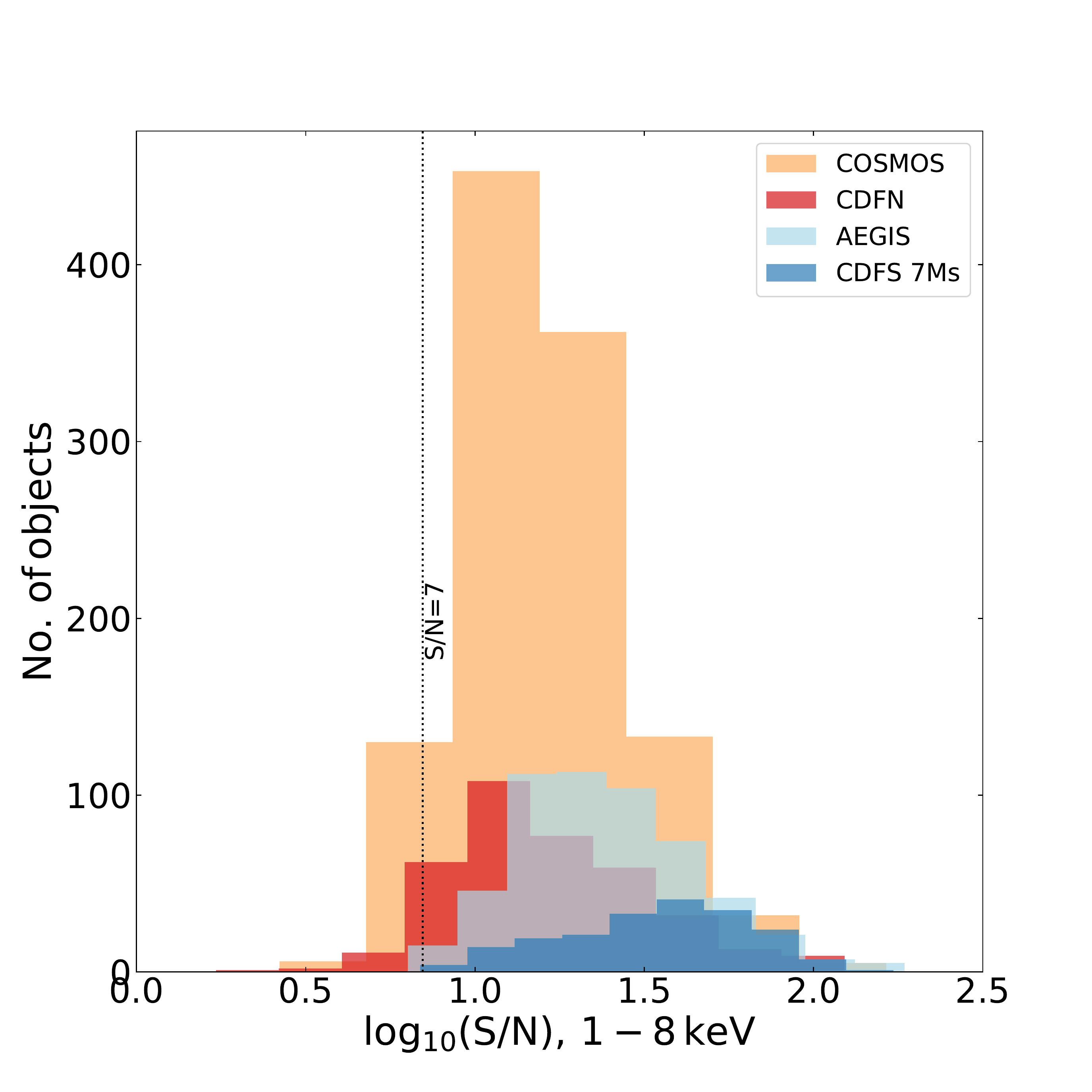}
\caption{\textit{Left}: Total counts (source counts plus background counts). \textit{Right}: Signal to noise ratio 
of the four samples COSMOS, AEGIS, CDFN and CDFS 7Ms.}
\label{fig:checkSNR}
\end{figure*}

\subsection{Data}

In this work we analyse four of the deepest fields observed by \textit{Chandra}, the 7Ms exposure of the \textit{Chandra} Deep Field South (CDFS), the \textit{Chandra} Deep Field North (CDFN), AEGIS and the COSMOS fields. The long exposures in these fields ensure that meaningful spectral information can be extracted for typical AGN at moderate redshifts ($z=0-4$) which dominate the accretion history of the Universe \citep[e.g.][]{Aird2010,Buchner2015}. In our analysis, we focus on the properties of the iron K$\alpha$ emission line and hard X-ray reflection continuum from Compton thick structures surrounding the AGN, such as the accretion disk and molecular torus. We hence use a hard X-ray selection ($>2$ keV) for all fields to ensure that meaningful spectral constraints can be obtained using these features. Here we summarise the data and sample selection in each field: 

 \subsubsection{The \textit{Chandra} Deep Field South}

The CDFS \citep{Luo2017} is, with a nominal total exposure time of $\sim$7 Ms, the deepest of all the \textit{Chandra} surveys, and indeed the deepest X-ray survey of all. While it reaches extremely faint fluxes, it covers a relatively small area of $\sim$0.13 deg$^2$ 
The CDF-S 7Ms is a collection of observations performed over multiple epochs between Oct 14, 1999 and Mar 24, 2016. 

All 102 observation used ACIS-I, which offers spectral imaging over an approximataely $17 \times 17$~arcmin  field of view 
and is often used for surveys. 

We limit our CDFS sample to the 199 hard X-ray ($2-7$ keV) selected AGN at redshift $\mathrm{z<4}$ previously studied in \citet{Baronchelli2018} and \citet{Buchner2014}. These were selected form the source catalog of \citet{Rangel2013}, which was based on the 4Ms \textit{Chandra} exposure. In the current analysis, however, we extract the spectra of the 4Ms sources from the deeper 7Ms exposure. 
The 7Ms spectra of these 199 sources contain a total of 313914 source counts in the observed 1--8 keV energy band, 
which is most relevant for our analysis. All of these sources have a redshift measurement, the majority of which are spectroscopic redshifts for which we adopt a single value. For the remaining 38$\%$ of these sources (76/199) photometric redshifts and their probability distributions from \citet{Hsu2014} were used. These are specially tailored for AGN, and are based on the methods of \citet{Salvato2009} and \citet{Salvato2011}.

\subsubsection{The \textit{Chandra} Deep Field North}

The CDFN is a field which is has received a \textit{Chandra} exposure of 2 Ms over a sky area of $\sim$0.12 deg$^2$
It comprises 20 different pointings taken between November 1999 and February 2002. We use the source catalog from \citet{Xue2016} to obtain the redshift values and to exclude stars. 
We analyse a sub-sample of 376 sources with redshift information from the 411 hard selected (2--7 keV band) sources in \citet{Xue2016}. Of these sources, 159 have spectroscopic redshift \citep{Xue2016}. 
The spectra contain a total of 
127563 source counts in the 1--8 keV energy band. The PDFs of the photometric redshift for the CDFN are not available, thus we used the preferred redshift adopted in \citet{Xue2016}. The fact that the photo-z PDFs are not available for the CDF-N means that, in some cases, an inaccurate redshift will be adopted in the spectral fit. 
On average, this will have 
the effect of reducing the significance when comparing the true underlying model with any other model, so should be conservative with respect to the significance
of the results presented below. 
Because a single value for the redshift is used, it will also result in an overly narrow posterior distributions for those sources.
This would also tend to lead to an overestimate of the intrinsic scatter and an 
underestimate of the statistical uncertainty of the mean values derived for the sample.

\subsubsection{AEGIS-X}

The AEGIS-X Deep survey \citep{Nandra2015} is the result of deep \textit{Chandra} imaging of the central region of the Extended Groth Strip. The survey encompasses an area of approximately 0.29 $\mathrm{deg^2}$ with a nominal exposure time of 800 ks. AEGIS-X is currently the third deepest \textit{Chandra} blank field survey after the \textit{Chandra} Deep Fields (CDF). While being shallower than the CDFs by a factor of $\sim$2--3 it covers an area $\sim3$ times larger. We use the source catalog from \citet{Nandra2015}, selecting as a parent sample the sources detected in the $2-7$~keV band, comprising 572 sources. After removing the sources identified as stars in \citet{Buchner2015}, we select a sub-sample of 540 sources with redshift information from the original 572 sources in the hard selected sample, with a total of 119462 source counts in the 1--8 keV energy range. Of these 540 sources, 
202 ($\sim 37\%$)  have a spectroscopic redshift. For the remainder, \citet{Nandra2015} provide photometric redshifts tailored for AGN, and their probability distribution functions, which we use in the spectral fitting.

\subsubsection{COSMOS}

The \textit{Chandra} COSMOS Legacy survey spans an area of 2.2 deg$^2$ on the sky. The central 1.5 deg$^2$ has a nominal exposure of $\mathrm{\sim 160}$ ks while the surrounding regions are nominally exposed with $\mathrm{\sim 80}$ ks depth \citep{Civano2016}. This makes it the largest area survey in our complication, but also the shallowest. We study the hard band selected sample from \citet{Civano2016} after removing the sources identified as stars in \citet{Buchner2015}, which comprises 1122 objects with a total of 95014 source counts in the 1-8 keV energy range. Of these sources, 534 have spectroscopic redshift, while the remaining 53$\%$ have accurate photometric redshifts and photo-z probability distributions from the work of \citet{Salvato2009}.

\subsection{Combined sample}

\begin{figure}
\centering
\includegraphics[width=\columnwidth,trim={1cm 0.5cm 2.2cm 2.7cm},clip]{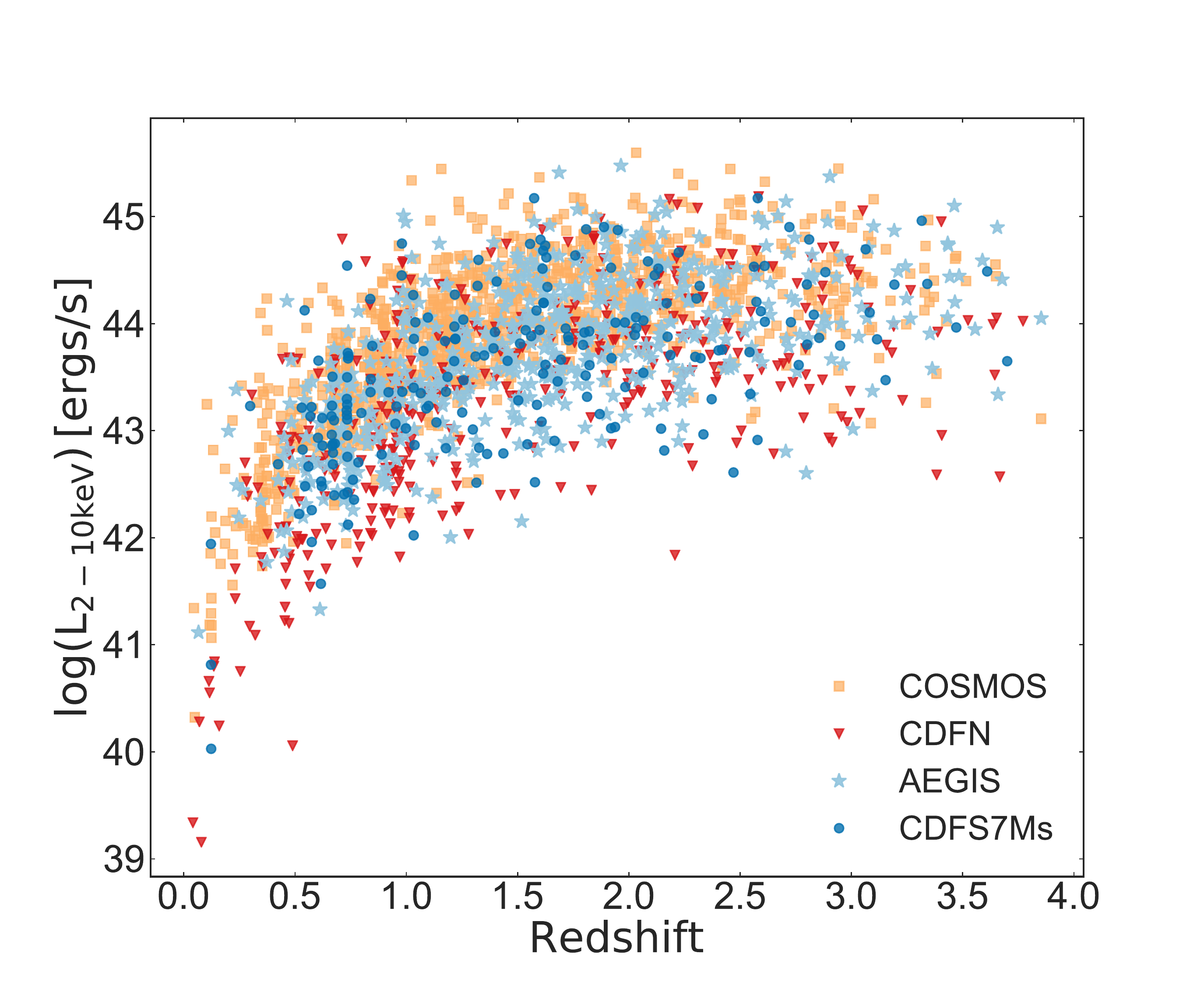}\vspace{1.2cm}
\includegraphics[width=\columnwidth,trim={1cm 0.5cm 2.2cm 2.7cm},clip]{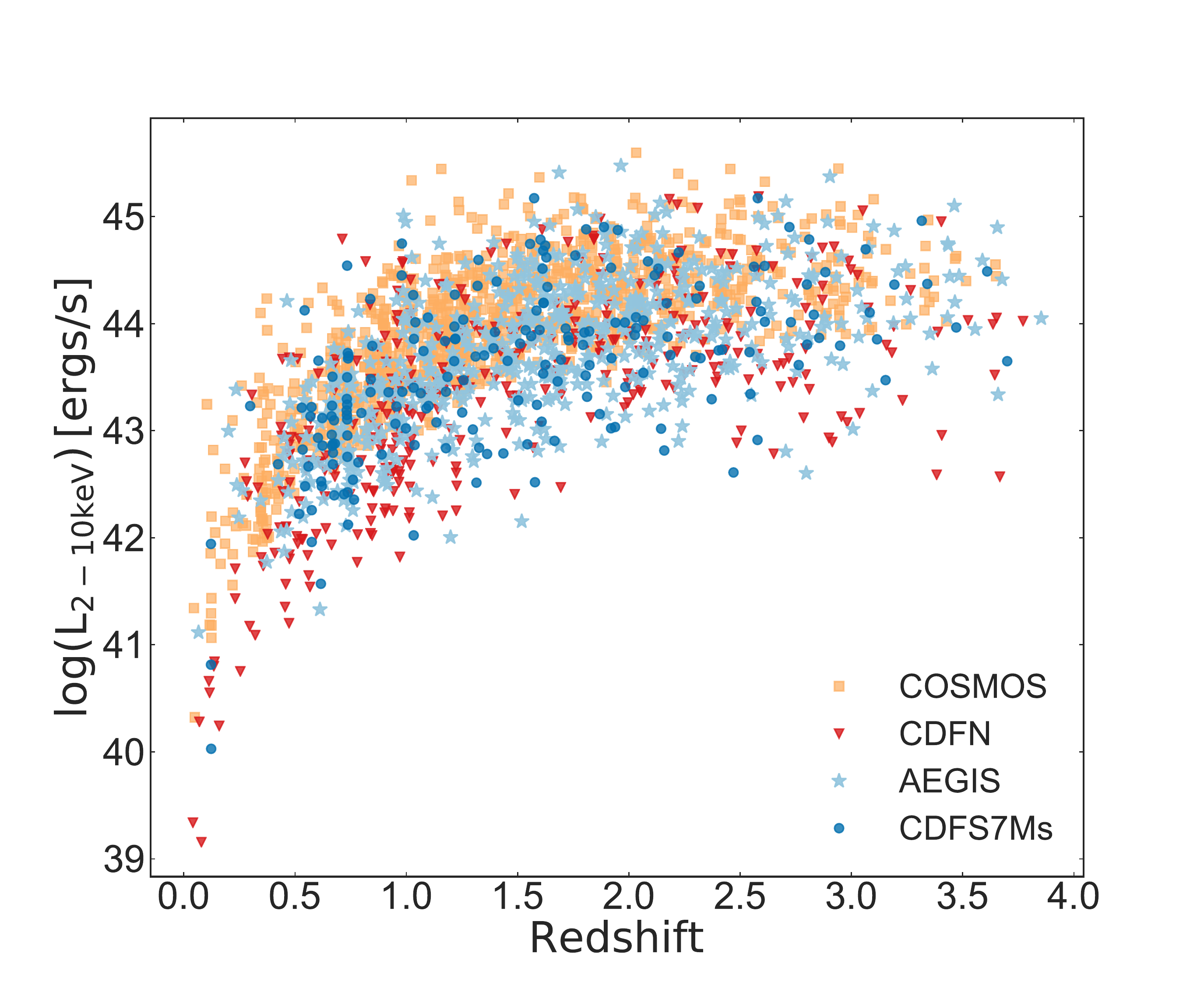}
\caption{2--10 keV Luminosity-redshift distribution for the full sample (\textit{Top}) and S/N > 7 subsample (\textit{Bottom}). We used the median of the absorption-corrected luminosity posterior probability from the model \texttt{zwabs*(zpowerlw + pexmon)}. The redshift used for the plots is the reported value from the original survey catalogs.}.

\label{fig:Lum}
\end{figure}

The parent sample for this study, combining all four fields, comprises a total of 2237 sources (See Table~\ref{SampleDetails}). 
Figure \ref{fig:checkSNR} shows the distribution of counts and S/N of our sample. 
We calculate the S/N using the formalism of \citet{Li1983} (see also \citet{Vianello2018}), which takes into account the Poisson nature of both the source and background count measurements. In particular we use the function \texttt{poisson$\_$poisson} from the python library \texttt{gv$\_$significance} developed by \citet{Vianello2018}. The sample spans a wide range of both total counts and signal-to-noise ratio. Numerically it is dominated by sources from the COSMOS survey, which has the largest area, but the objects from the deepest field, the CDF-S, have the highest number of counts and S/N overall. The right hand panel of  Fig. \ref{fig:checkSNR} shows a vertical line at S/N=7. We use sub-samples cut at this S/N in the subsequent analysis, as discussed below, and the number of sources and source counts obtained after applying this S/N cut are shown in Table~\ref{SampleDetails}. 

The luminosities $\mathrm{L}$, defined in the 2--10 keV energy range, and redshifts of the selected sources are presented in Figure \ref{fig:Lum}, which also shows the S/N split sub-samples. The luminosities were calculated from the spectral fits described below and corrected for galactic and intrinsic absorption. 
The sample covers a broad redshift range up to $z\sim 4$, and the bulk of the sample covers the luminosity range $\log(L/\mathrm{erg\,s^{-1}})=42-45$\footnote{The log here represents the logarithm in base 10 ($\mathrm{\log_{10}}$).}, with just a very few low redshift sources fainter than this. The sample becomes increasingly incomplete at luminosities below $\log(L/\mathrm{erg\,s^{-1}})=43$ above a redshift of about 1.   

\subsection{Spectral Extraction}
\label{specex}

We extracted the spectra of the sources using the software package \textit{ACIS Extract} (AE) \citep{Broos2010,Broos2012}.
AE was developed to automate as much as possible the analysis of X-ray data taken with the \textit{ACIS} instrument of \textit{Chandra}. It is well suited for our application as, given a source catalog, spectra can be extracted from multiple observations of the same field. As input files, AE requires the Level 2 event list of the observations, the exposure maps and aspect histograms corresponding to the field of view of the event data, the aspect solution file covering the time range of the observations and the mask file of the observations. We produced these files with a combination of the \textsc{ciao 4.10} \citep{Fruscione2006} and \textsc{ftools 6.25} \citep{Blackburn1995} software. 

The process of extracting spectra using AE can be summarized in four steps. First, the AE tool \textit{ae$\_$make$\_$catalog} builds extraction regions sized to encompass 90$\%$ of the local point spread function (PSF) but small enough to avoid overlaps in crowded regions.
Secondly, the tool \textit{ae$\_$standard$\_$extraction} extracts source and background spectra of the sources in the catalog. Note that for data taken at -110C on certain CCDs the event file will not be corrected for charge transfer inefficiency (CTI) and AE will be set to use the \textsc{ciao} tool \textit{mkrmf} instead of the default \textit{mkacisrmf} to build the RMF files. The next step uses \textit{ae$\_$adjust$\_$backscal$\_$range} to analyze the source's existing background extraction region and to choose a target background scaling range individually for each source. The process of extracting background and choosing a scaling range has to be repeated until the scaling range is stable. Finally, AE merges the observations combining the extraction from all the ObsIds and performing the photometry.

We produced the appropriate input files for AE following the method presented in \citep{Georgakakis2011}. The final data products are the source and background spectra together with RMF and ARF files.

\begin{table}
\caption{Parameter description for model \texttt{zwabs*(zpowerlw + pexmon + kerrconv(pexmon))} 
\label{eq:blur}. The model component blur represents the model kerrconv. We fix the spin parameter to the values 0 or 0.998 depending on the case we want to analyze. The model has six free parameters: the column density $\mathrm{N_H}$, the photon index, the inclination of the broad component and the three norms. The strength of the blurred reflection component component $R_{blur}$ is measured relative to the power law and is defined as the ratio of the normalization of the blurred pexmon component ($A_{blur}$) to that of the power-law ($A_{pow}$). The parameters that have units are [$N_H$] = \textbf{atoms $\mathrm{cm^{-2}}$}, [foldE] = keV, [Incl] = deg, $\mathrm{r_{br}}$ in gravitational radii and $\mathrm{R_{in}}$ and $\mathrm{R_{out}}$ in units of the radius of marginal stability.}
\begin{tabular}{lllllll}
\hline \hline
Comp.$^{a}$ & \multicolumn{1}{l}{No.$^{b}$} & Name$^{c}$ & Min & Max & Fix val. & Free \\ \hline
zwabs & 1 & log($\mathrm{N_H}$) & 20 & 26 & - & yes \\ 
 & 2 & Redshift & - & - & \multicolumn{1}{l}{z} & - \\ 
zpowerlw & 3 & PhoIndex & \multicolumn{1}{l}{1.1} & \multicolumn{1}{l}{2.5} & - & yes \\ 
 & 4 & Redshift & - & - & \multicolumn{1}{l}{link to 2} & - \\ 
 & 5 & $\log \mathrm{A_{pow}}$ & -10 & 1 & - & yes \\ 
pexmon & 6 & PhoIndex & - & - & link to 3 &  -\\ 
 & 7 & foldE & -& - & 800 & - \\ 
 & 8 & rel$\_$refl & - & - & \multicolumn{1}{l}{-1} & - \\ 
 & 9 & redshift & - & - & \multicolumn{1}{l}{link to 2} & - \\ 
 & 10 & abund & - & - & \multicolumn{1}{l}{1} & - \\ 
 & 11 & Fe$\_$abund & - & - & \multicolumn{1}{l}{1} & - \\ 
 & 12 & Incl & - & - & \multicolumn{1}{l}{60} & - \\ 
 & 13 & $\mathrm{log(R_{pex})}$ & -2 & 1 & $\log \mathrm{\frac{A_{pex}}{A_{pow}}}$ & yes \\ 
kerrconv & 14 & Index1 & - & - & \multicolumn{1}{l}{3} & - \\ 
 & 15 & Index2 & - & - & 3 & - \\ 
 & 16 & $r_{br}$ & - & - & 6 & - \\ 
 & 17 & Rin & - & - & 1 & - \\ 
 & 18 & Rout & - & - & 400 & - \\ 
 & 19 & Spin & - & - & 0/0.998 & -\\
 & 20 & cos(Incl) & 0 & 1 & - & yes \\ 
pexmon & 21 & PhoIndex & - & - & link to 3 & - \\ 
 & 22 & foldE & - & - & 1000 & - \\ 
 & 23 & rel$\_$refl & - & - & -1 & - \\ 
 & 24 & redshift & - & - & \multicolumn{1}{l}{link to 2} & - \\ 
 & 25 & abund & - & - & 1 & - \\ 
 & 26 & Fe$\_$abund & - & - & 1 & - \\ 
 & 27 & Incl & - & - & link to 17 & - \\ 
 & 28 & $\mathrm{log(R_{blur})}$ & -2 & 1  & $\log \mathrm{\frac{A_{blur}}{A_{pow}}}$ & yes \\ \hline \hline
\end{tabular}
\label{blur}

\flushleft 
\footnotesize{$^{a}$Model component.}\\
\footnotesize{$^{b}$Parameter number.}\\
\footnotesize{$^{c}$Parameter name.}\\
\end{table}


\subsection{Model fitting and model comparison}
\label{models}

Four physically justified models are considered to represent different scenarios for the major gas structures surrounding the central black hole. The first model is an absorbed power-law, \texttt{zwabs*zpowerlw} in \textsc{Xspec} terminology, which describes the emission from an X-ray corona behind a screen of obscuring gas. In the second model we add a non-relativistic reflection component to the simple absorbed power-law, \texttt{zwabs*(zpowerlaw+pexmon)}, to represent reflection from distant material such as the obscuring torus. We chose the \texttt{pexmon} model \citep{Nandra2007} to describe the reflection component since it combines a exponentially cutoff power-law emission reflected by neutral material (\texttt{pexrav}; \citet{Magdziarz1995}) with self-consistently generated Fe K$\alpha$, Fe K$\beta$, Ni K$\alpha$ and Fe K$\alpha$ Compton shoulder emission \citep{George1991,Matt2002}. The third and the fourth model add a relativistically broadened reflection component to the second model, as expected from an accretion disk. This last component is modelled by convolving a narrow reflection spectrum with a convolution model to represent the expected Doppler and gravitational shifts expected from an accretion disk. Specifically, we use the the \texttt{kerrconv(pexmon)} model \citep{Brenneman2006}. The \texttt{kerrconv} allows the BH spin to be a free parameter. However, to avoid having unnecessarily many free parameters in the broadened reflection model we constrain our analysis by fixing the spin parameter to the two special cases of spin $a=0$ (Schwarzschild metric, model three) and maximally spinning $a=0.998$ (model four). The parameter priors are chosen to be consistent with \citet{Baronchelli2018}, except for the parameter describing the inclination angle of the broad component, that in the current work is chosen to be uniformly distributed in cosine space (see Table \ref{blur}). In Table \ref{blur} we list the chosen model parameters for the components \texttt{xszwabs}, \texttt{powerlaw}, \texttt{pexmon} and \texttt{kerrconv(pexmon)}. For the parameters allowed to vary, we list the minimum and maximum value of the prior distributions. The normalization parameters and the $N_H$ are chosen to be uniform in logarithmic space, the inclination angle in \texttt{kerrconv} is uniform in cosine space, while all the other parameters are uniform in linear space. When only the photometric redshift of the source is available, we use the probability distribution function (PDF) produced using the SED fitting procedures and templates from \citet{Salvato2009} and \citet{Salvato2011} to take into account of the uncertainty of the photometric redshift estimation. The exception is the CDFN for which the photometric redshift PDFs are not available, and we use a single value for the redshift.

We fit a background model simultaneously to the data, following \citet{Buchner2014}, instead of subtracting the background. We fit the \textit{Chandra} background model provided in BXA to all the background spectra and then we include the background model with best fit parameters frozen in the model of the source spectra \citep[see also][]{Buchner2014,Baronchelli2018}.
As discussed in the introduction, we do not stack the spectra to determine the average properties of the sample, but instead fit each source individually and then infer the properties of the sample by combining the information from these individual fits.

\begin{table*}
\caption{Comparison of the Bayesian evidence for our four models for the CDFS 7Ms, CDFN, AEGIS, COSMOS and the full sample. In each column, the values of $\mathrm{\log(Z)}$ for each model are normalized by the $\mathrm{\log(Z)}$ of the model with highest evidence. Thus in this table, the model with highest evidence is identified by a value of $\mathrm{\log(Z)}=0$. We fit the models in the observed frame energy range 1 -- 8 keV.}
\begin{tabular}{llccccc}
\hline \hline
& Model $^{a}$& Total & CDFS 7Ms & CDFN & AEGIS & COSMOS \\
 & & \multicolumn{1}{r}{log(Z)$^{b}$} & \multicolumn{1}{r}{log(Z)$^{b}$} & \multicolumn{1}{r}{log(Z)$^{b}$} & \multicolumn{1}{r}{log(Z)$^{b}$} & \multicolumn{1}{r}{log(Z)$^{b}$} \\
 \hline
 \textbf{S/N $\geq$ 0}  &  &  & &  &  &    \\ 
 \hline
& \texttt{zwabs*(zpowerlw)} & -438.6 & -133.2 & -103.8 & -101.0 & -110.1\\ 
& \texttt{zwabs*(zpowerlw+pexmon)} & -26.8 & -10.2 & -19.7 & -7.2 & 0\\ 
& \texttt{zwabs*(zpowerlw+pexmon+kerrconv0(pexmon))}& -6.3 & -2.9 & -2.3 & -3.1 & -10.6 \\ 
& \texttt{zwabs*(zpowerlw+pexmon+kerrconv1(pexmon))}&  0 & 0 & 0 & 0 & -8.8\\ 
 \hline
 \textbf{S/N $\geq$ 7}  &  &  & &  &  &    \\ 
 \hline
& \texttt{zwabs*(zpowerlw)} &  -437.1 & -133.6 & -102.1 & -101.0 & -109.5\\ 
& \texttt{zwabs*(zpowerlw+pexmon)} &  -27.0 & -10.4 & -19.2 & -7.2 & 0\\ 
& \texttt{zwabs*(zpowerlw+pexmon+kerrconv0(pexmon))}&  -6.6 & -3.0 & -2.5 & -3.1 & -10.0 \\ 
& \texttt{zwabs*(zpowerlw+pexmon+kerrconv1(pexmon))}&  0 & 0 & 0 & 0 & -8.2\\ 
 \hline
 \hline
\end{tabular}
\label{CDFS7Ms}
\flushleft 
\footnotesize{$^{a}$ Model components.}\\
\footnotesize{$^{b}$ Logarithm of the Bayes evidence of the full sample normalized to the largest evidence.}\\

\end{table*}

\subsection{Bayesian X-ray Analysis}

The \textit{Bayesian X-ray Analysis} (BXA) \citep{Buchner2014} package is a Bayesian framework to determine the best fit parameters and their posterior distribution for X-ray spectra. 
BXA applies \texttt{PyMultinest} \citep{Buchner2014}, a python wrapping of an implementation of the nested sampling algorithm (\textsc{Multinest}) \citep{Feroz2008} combined with \textsc{Sherpa} or \textsc{Xspec} to compute the  Bayesian evidence Z for X-ray data and hence parameter constraints. The Bayesian evidence Z is the integral of the likelihood over the prior and can be interpreted as the probability $P(D|M)$ of the model $M$ given the data $D$ marginalized over the model parameters $\theta$.

\begin{equation}
    Z=P(D|M)=\int L(\theta)P(\theta|M)d\theta,
\end{equation}
\textsc{MultiNest} \citet{Feroz2008,Feroz2009,Feroz2013} provides an efficient approximation to this integral. This algorithm samples a number of live points in the parameter prior space evaluating the likelihood of the model for every point. At every step, the live point with lowest likelihood will be replaced by a newly sampled point until the algorithm converges to the highest likelihood value. \textsc{MultiNest} is particularly specialized to deal efficiently with multi-modal distributions by using a recursive clustering algorithm and proposal regions in the shape of ellipsoids.

To perform model comparison, we assume that all sources are described by the same model. We calculate the total evidence for that model by adding the $\log(Z)$ \citep[see also, ][]{Buchner2014,Baronchelli2018} values from the individual fits. This allows one to compare the total evidence of the sample for different models. The difference in the logarithmic evidence then corresponds to a Bayes factor (BF), which can be used to discriminate between the models. A commonly used way to interpret the BF values is the Jeffrey scale, which strengthens the choice of one model over the other approximately every time that the logarithm of the BF increases by one in natural logarithmic units \citep{Robert2009}.  However, Bayes factors are continuous quantities and such discretisations should not be over-interpreted \citep{Nesseris2013}. 
In \citep{Baronchelli2018}, we verified with simulations that the Bayes factors scatter around one ($\Delta \log Z\approx 0$) for low signal-to-noise data.

\section{Results}
\label{results}

\begin{figure}
\centering
\includegraphics[width=\columnwidth,trim={0 0 2cm 2.7cm},clip]{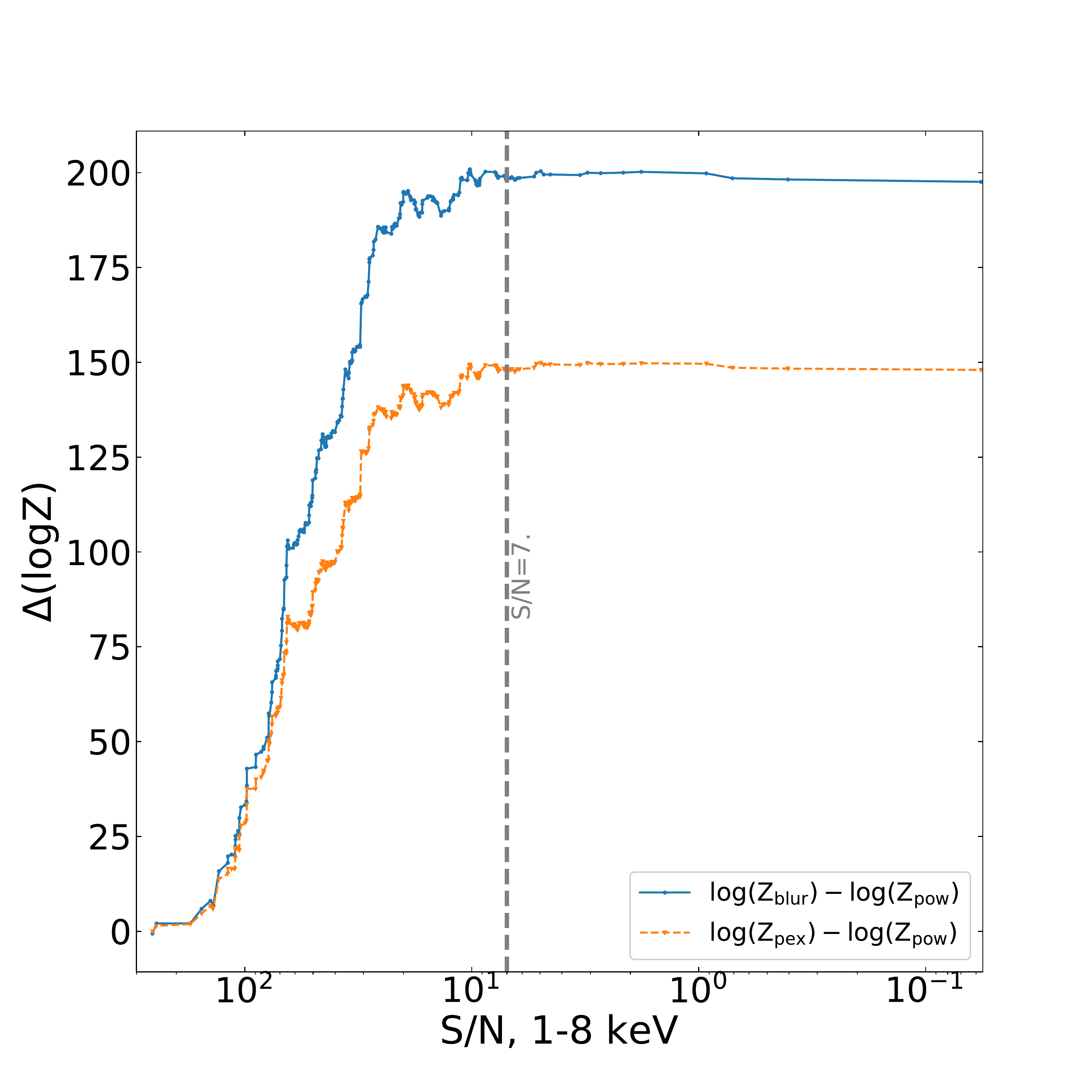}
\includegraphics[width=\columnwidth]{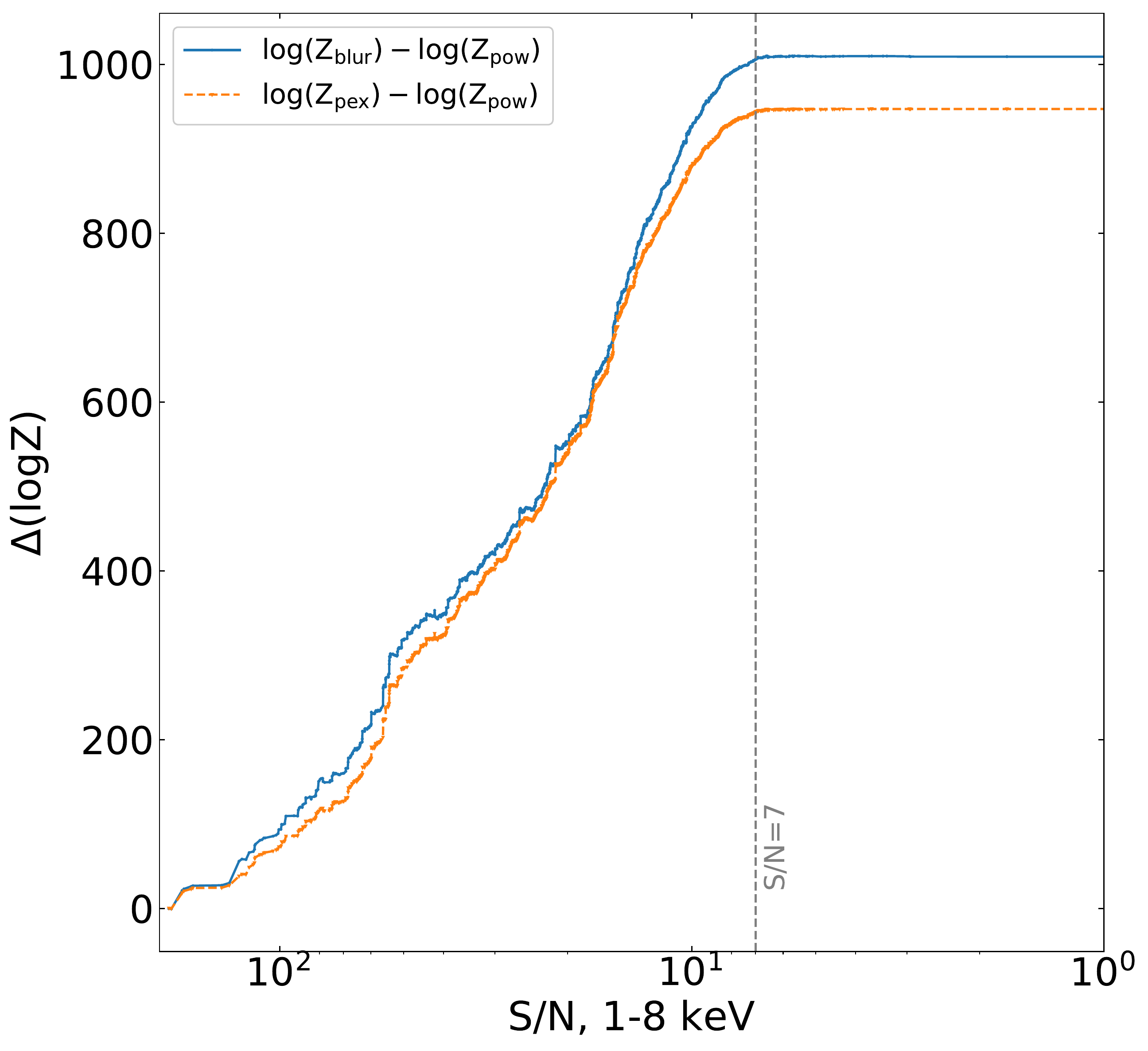}

\caption{Cumulative evidence of the full sample for decreasing S/N for the full sample (\textit{Bottom}) and the simulated sample (\textit{Top}). The circular points (blue, solid line) show the cumulative evidence for the blurred model normalized by the evidence of the simple power-law. The triangles (orange, dashed line) show the cumulative evidence for the narrow model normalized by the evidence of the simple power-law. For the full sample the blue curve remains above the green curve. The gray dashed vertical line at $\mathrm{S/N=7}$ shows the S/N limit below which adding the evidence of the single sources to the total value does not add further information.}
\label{fig:cumulZ}
\end{figure}

\subsection{Initial fitting and S/N effects}
\label{simulations}

We first fit the total sample of 2237 sources with BXA to calculate the Bayesian evidence Z for the four models described in Section \ref{models}. We fit each source with each of the models individually and use the Bayes factor (BF) method to compare the models and determine which one provides the better fit. The single sources are generally too faint to significantly favour one model over the others. Instead, as discussed above, we combine the Bayesian evidence from the individual source fits to obtain the evidence for the full sample \citep{Buchner2014,Baronchelli2018}. 

As can be seen in Table \ref{CDFS7Ms}, the Bayesian evidence for the total sample shows a strong preference for the model including both narrow and broad reflection components, with a maximally spinning black hole preferred over a Schwarzschild solution. This preference is also found in the field-by-field subsamples, with the exception of the COSMOS field. Here the preferred model is that without the blurred reflection component. We discuss this result further below, but note here that there are a number of differences between the COSMOS sample and the remainder of the fields. One such difference is that, as shown in Fig \ref{fig:checkSNR}, the COSMOS sources typically have lower signal-to-noise ratio than the sources in the other fields. This raises the possibility that, below a certain S/N threshold, the spectra become insensitive to the broad reflection component and do not add information about the presence or properties of that component. 

This hypothesis is borne out by the data. Fig. \ref{fig:cumulZ} (top) shows the cumulative evidence as a function of signal-to-noise ratio for the sample. The evidence rises rapidly when adding sources with high S/N, but then flattens off and eventually becomes approximately horizontal, showing that the lowest quality spectra are not adding additional information/evidence. To estimate at which S/N the source will not add further information to the total evidence of the sample, we perform a set of simulations using the \texttt{fakeit} tool of \textsc{Xspec} \citep{Arnaud1996}.

We simulate a sample of 300 sources using the ancillary files (ARF and RMF), the background spectra and the redshifts from the AEGIS sample. As a first step to simulate a sample of spectra, we have to choose an input model that will define the spectral shape of the simulations. Since we are interested in studying how our method would perform on a sample of relativistic broadened spectra at different S/N, we chose to simulate the spectra using the fourth model.  We simulate the sources so that the total number of net counts is comparable with the one of CDFS 7Ms sample.

We fit the simulated sample with the four models described in Section \ref{specex}. In the bottom panel of Fig. \ref{fig:cumulZ}, we show the cumulative evidence ratios $\log(Z_{blur})-\log(Z_{pow})$ and $\log(Z_{pex})-\log(Z_{pow})$ that represent the cumulative distributions of the broad and narrow models normalized by the evidence of the simple powerlaw. It can be seen that the two curves start to flatten below S/N$\mathrm{\sim 7}$, very similar to what is seen with the real data. We thus conclude that below this S/N value, little additional information is being added about the properties of the reflection, and henceforth restrict our analysis to a subsample with S/N$>7$. The number of sources meeting this S/N criterion are shown in Table~\ref{SampleDetails}.

\subsection{Compton reflection properties of the sample}

\begin{table}
\caption{Same as Table \ref{CDFS7Ms} but adding a model, \texttt{zwabs*(zpowerlw+kerrconv(pexmon))},  with a broad reflection component and no narrow reflection to the comparison for the COSMOS field. In each column, the values of $\mathrm{\log(Z)}$ for each model are normalized by the $\mathrm{\log(Z)}$ of the model with highest evidence. Thus in this table, the model with highest evidence is identified by a value of $\mathrm{\log(Z)}=0$. We fit the models in the observed frame energy range 1 -- 8 keV.}
\begin{tabular}{lc}
\hline \hline
Sample/Model $^{a}$& COSMOS \\
& \multicolumn{1}{r}{log(Z)$^{b}$} \\
 \hline
 \textbf{All} &    \\ 
 \hline
 \texttt{zwabs*(zpowerlw+kerrconv1(pexmon))}& -2.1 \\ 
 \texttt{zwabs*(zpowerlw+pexmon)} & -1.7\\ 
 \texttt{zwabs*(zpowerlw+kerrconv0(pexmon))} & 0\\ 
 \hline
 \textbf{S/N $\geq$ 7} &    \\ 
 \hline
 \texttt{zwabs*(zpowerlw+kerrconv1(pexmon))}& -2.3 \\ 
 \texttt{zwabs*(zpowerlw+pexmon)} & -2.2\\ 
 \texttt{zwabs*(zpowerlw+kerrconv0(pexmon))} & 0\\ 
 \hline
 \hline

\end{tabular}
\label{COSMOS}
\flushleft 
\footnotesize{$^{a}$ Model components.}\\
\footnotesize{$^{b}$ Logarithm of the Bayes evidence of the full sample normalized to the largest evidence.}\\

\end{table}

The results of the model comparison for the S/N-censored sample are also given are given in Table \ref{CDFS7Ms}. The results are, in fact, rather similar to the full sample, as the number of very low S/N ratio sources is small. More specifically, we find that the BF method selects the model including both broad and narrow reflection components to be the best fitting model in the CDFS 7Ms, CDFN and AEGIS fields. 
For the COSMOS field, the BF method selects the model with only distant reflection as the best-fitting model over the more complex model with both narrow and broad reflection. 
Considering the total sample, comprising all four fields, the evidence for the broad reflection is very strong. The difference in the logarithmic evidence can be interpreted like a probability difference. Thus, we can see from Table \ref{CDFS7Ms} that the model containing a blurred component is selected to be $\mathrm{10^{26.8}}$ more probable than the scenario with only a narrow reflection component to describe the sample with S/N larger than 7.
The total evidence also shows that the model with a maximally spinning (Kerr) black hole has a probability $\mathrm{10^{6.6}}$ of being preferred over a non-rotating (Schwarzschild) solution, again seen also in the individual fields with the exception of COSMOS. 

The fact that the COSMOS field shows a preference for narrow reflection only in the evidence comparison shown in Table~\ref{CDFS7Ms} does not necessarily imply that broad reflection is not present in the COSMOS source population. 
This is because we test for the presence of the broad reflection {\it in addition} to narrower reflection from more distant material, e.g. the torus. As the model with broad reflection has more free parameters, this additional complexity is penalised in the evidence comparison. We therefore performed an additional test by fitting the spectra also with a model with a blurred reflection component but no additional narrow reflection component (see Table \ref{COSMOS}). This then tests whether there is evidence for broad reflection as an alternative to the narrow reflection. According to this test the preferred model is that with a blurred reflection component with a non-rotating (Schwarzschild) black hole. The preference for this model over the narrow reflection or maximally-spinning black hole is, however, marginal. We conclude that, while the COSMOS data are of sufficient quality to confirm the presence of Compton reflection in the spectra, they are not able to distinguish the properties of the reflection e.g. whether it is broad or narrow, or if broad the value of the black hole spin implied. Based on the fields with higher S/N ratio spectra, however, it seems most likely that both components are present also in the COSMOS data. 

\begin{figure*}
\centering
\includegraphics[scale=0.5]{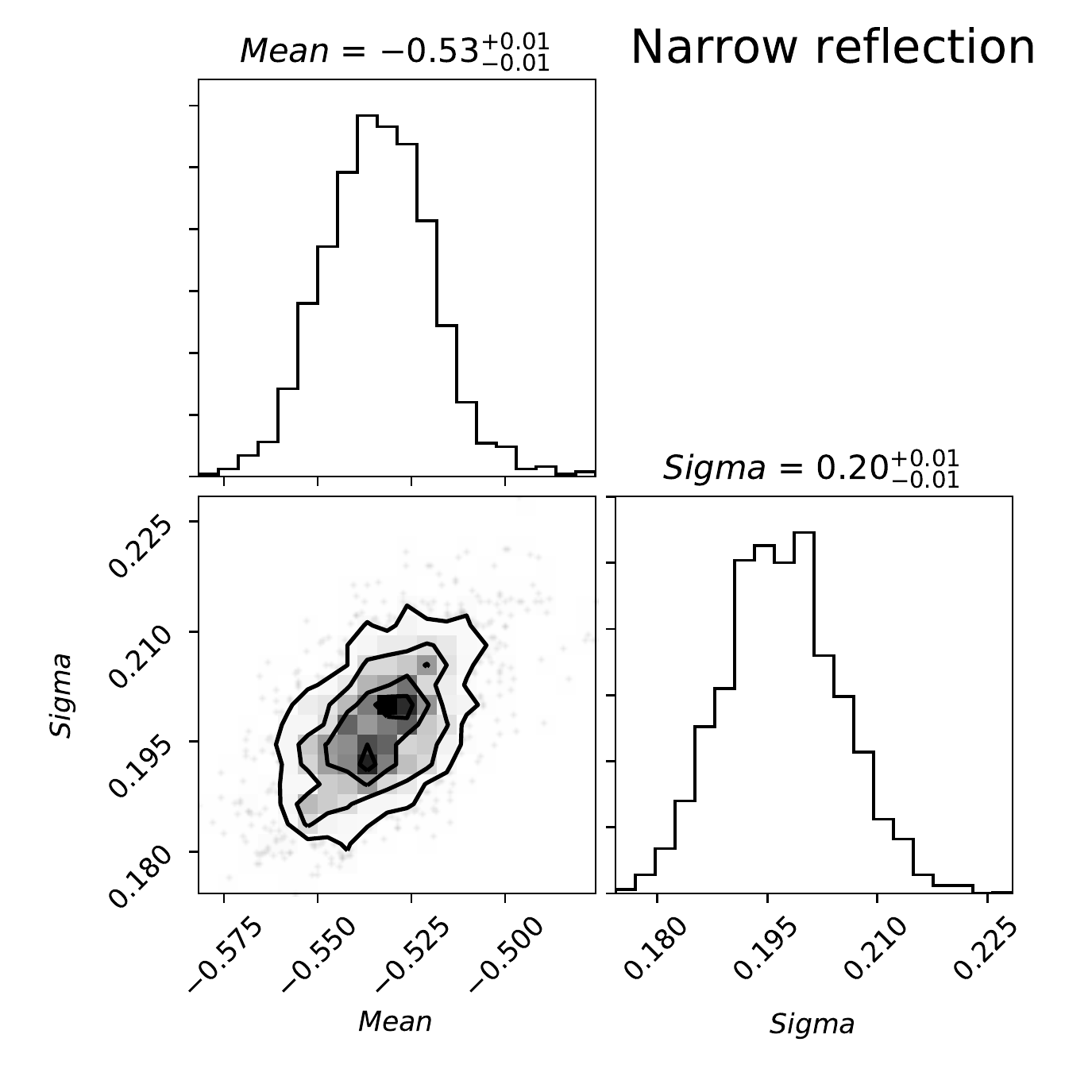}
\includegraphics[scale=0.5]{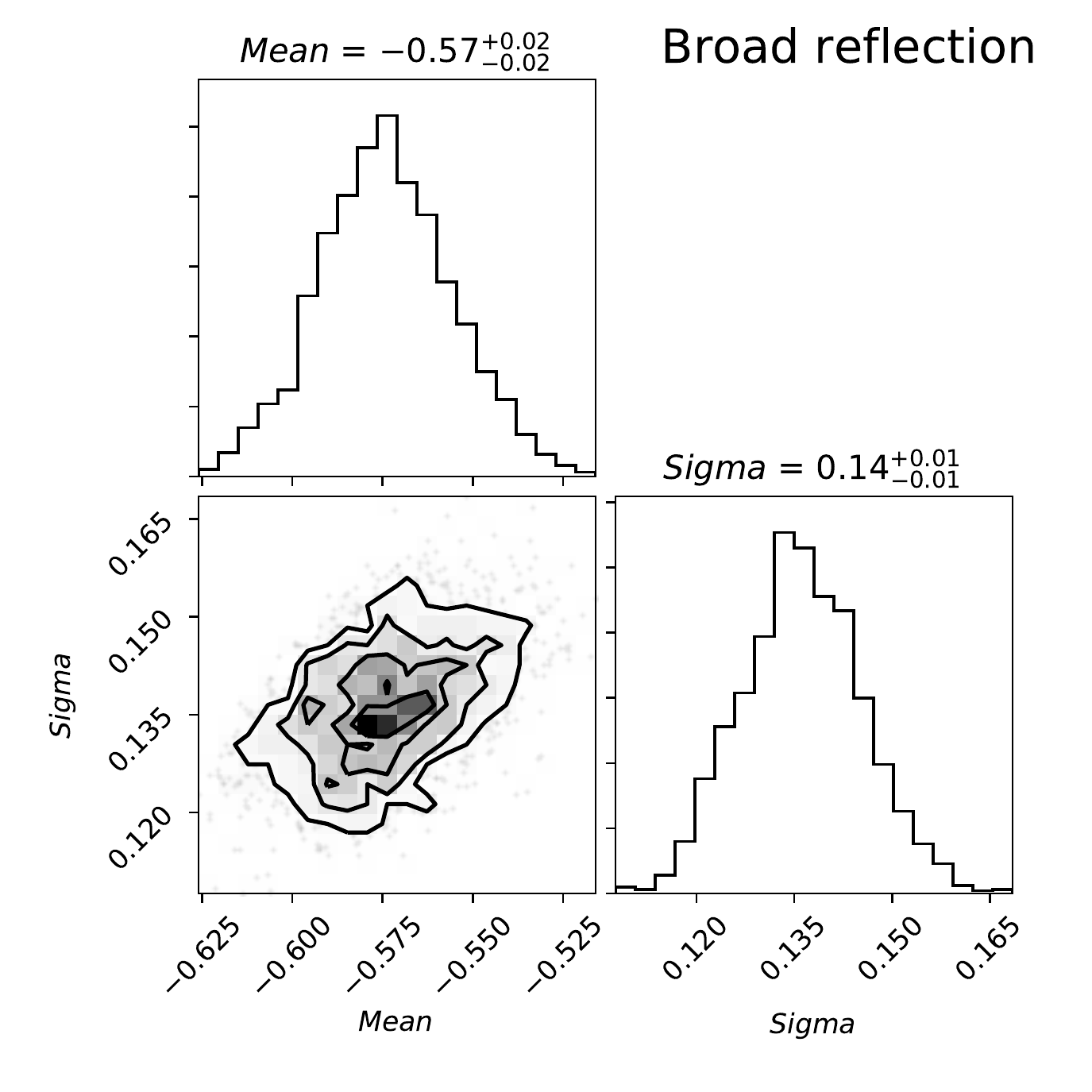}
\caption{Corner plots of the mean and sigma of the population of $\mathrm{log(R)}$ values for narrow (\textit{left}) and broad (\textit{right}) reflection component. These result were calculated using the method explained in Appendix \ref{appsec:stan}.}
\label{fig:corner}
\end{figure*}

The average strength of both the narrow and broad reflection components, as measured by the $R$ parameter, is an important diagnostic of the system, as it depends on the geometry and, in the case of the broad reflection, potentially also on relativistic effects close to the black hole \citep[e.g.][]{Miniutti2004}. The task of calculating the mean and sigma of the underlying parent population that describes the $R$ value is not trivial, however, since the posterior distributions for this parameter are not always well described by a normal distribution. To address this, we us a Hierarchical Bayesian model (HBM), described in detail in Appendix \ref{HBM}. For a sample it models the intrinsic log(R) distribution of the sample as a Gaussian. Taking into account the posterior uncertainties on each individual object, the HBM fit returns mean and standard deviation $\sigma$ of the distribution. The results are shown in Fig \ref{fig:corner}. Applying the HBM to sources with S/N>7 from all fields, we find a mean of $\mathrm{log(R_{pex}) = -0.53}$ (thus, $R_{\rm pex} = 10^{-0.53}=0.30$) with spread $\mathrm{\sigma_{log(R_{pex})} = 0.2}$ for the narrow reflection component and mean of $\mathrm{log(R_{blur}) = -0.57}$ (thus, $R_{\rm blur} = 10^{-0.57} = 0.27$) with spread of $\mathrm{\sigma_{log(R_{blur})} = 0.14}$ for the blurred reflection component. The population mean values obtained with the HBM for the reflection fraction are similar for the narrow and broad components. The strength of the blurred reflection component in particular is smaller than would be expected from a flat disk illuminated by a point source, as has been found previously \citep{Nandra2007}. Both $R$ values also show a significant spread of $\sim 0.2$ dex, indicating there is considerable diversity in reflection strength within the population.

\begin{figure*}
\includegraphics[scale=0.4]{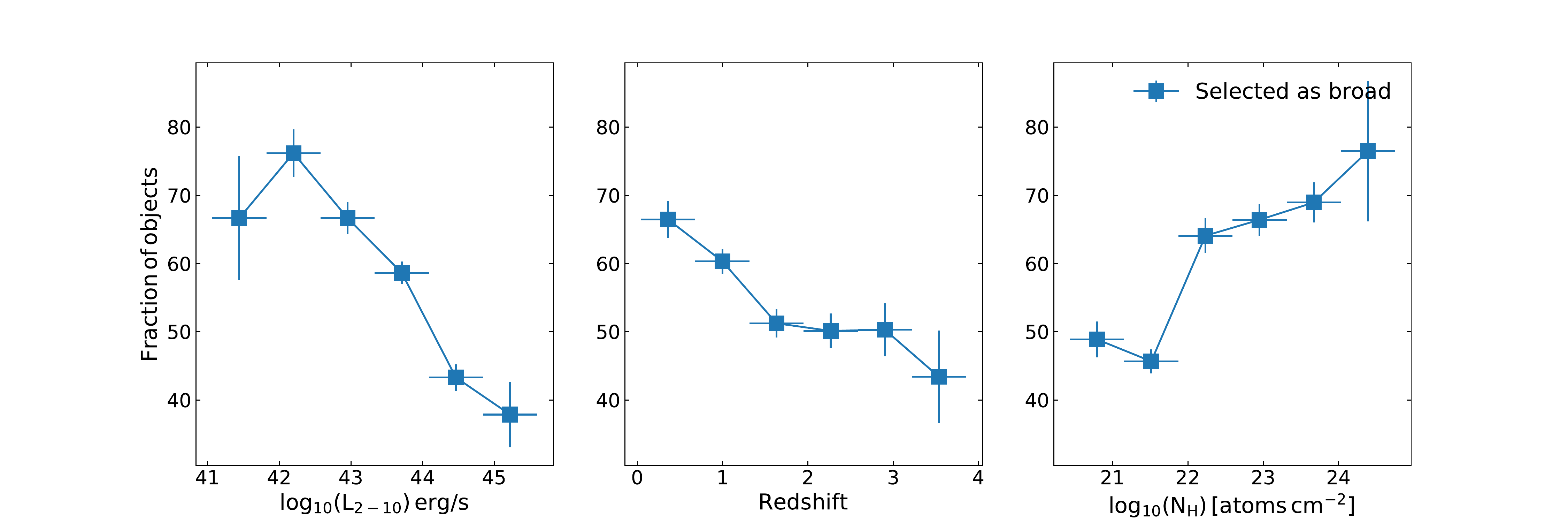}
\caption{Fraction of sources in the sample with S/N$\geq$7 best fitted by a model with broad reflection component as a function of luminosity (\textit{right}), redshift (\textit{middle}) and column density $\mathrm{N_H}$ (\textit{left}).
The fraction decreases with higher luminosities hinting that an anti-correlation with the Fe K$\alpha$ line EW and the intensity of the luminosity might be present.}
\label{fig:Baldwin}
\end{figure*}

\begin{figure*}

     \subfloat{%
        \includegraphics[scale=0.38]{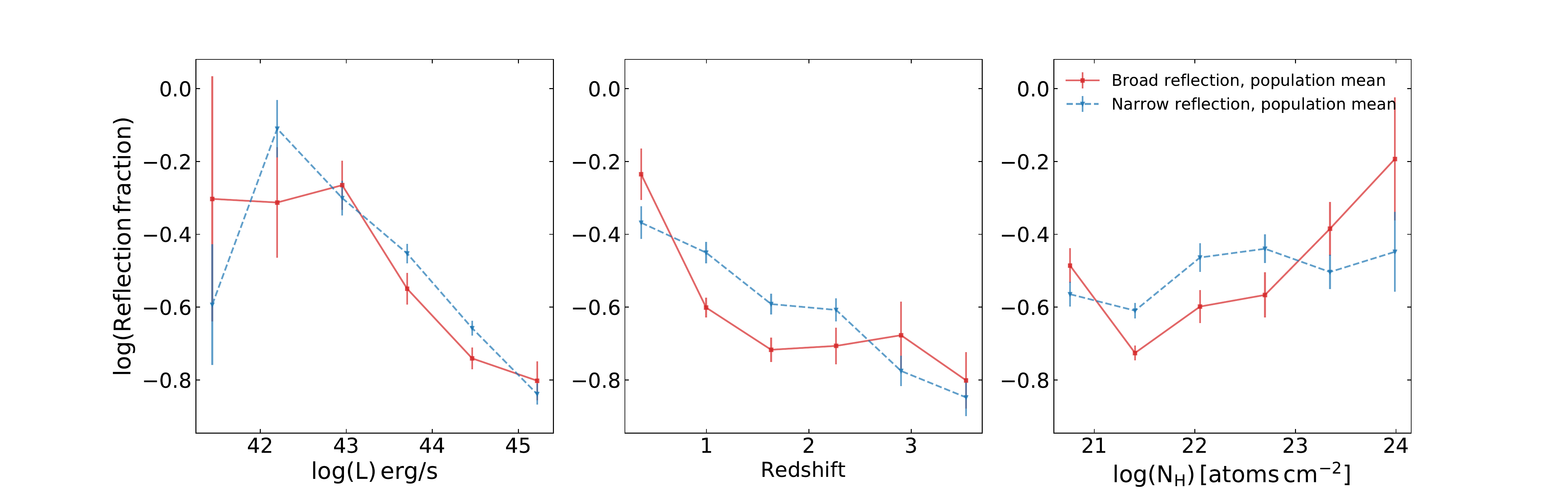}
    }

     \subfloat{%
        \includegraphics[scale=0.38]{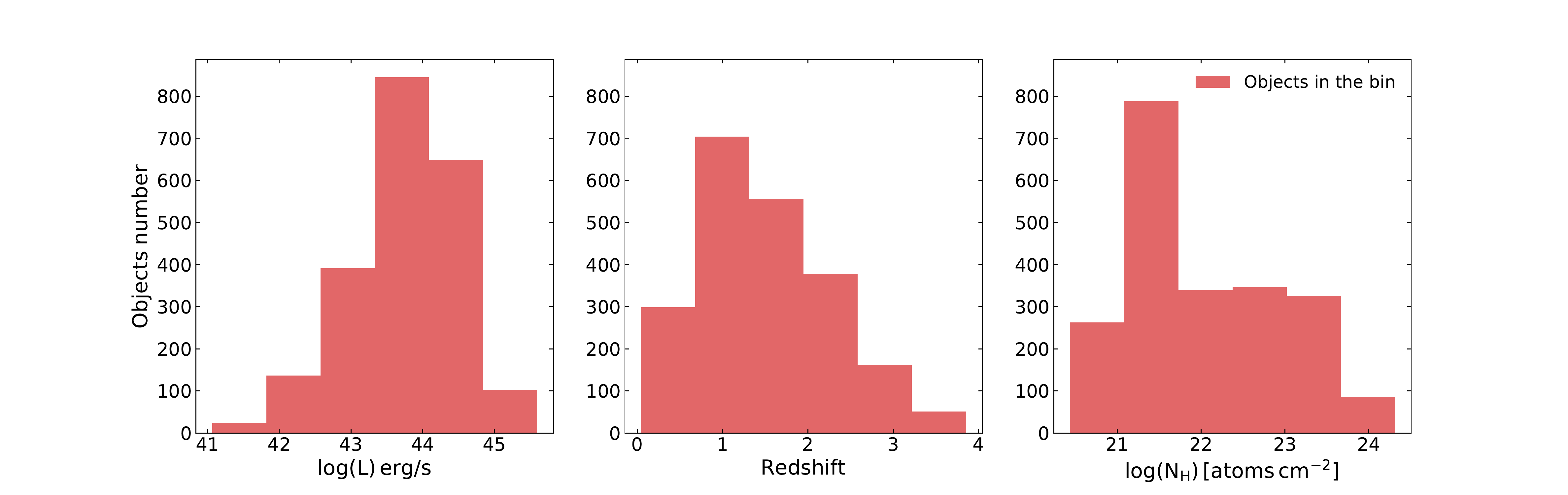}
    }
    \caption{Dependence of the reflection fraction for the broad accretion disk reflection component $R_{\rm blur}$ (red) and the narrow reflection $R_{\rm pex}$ from distant material (blue) as a function of the luminosity (\textit{Left panels}) redshift (\textit{Middle panels}) and obscuring column density (\textit{right panels}). Sources with $\log L<41$ erg/s were excluded from the analysis.
    The upper panels show the mean and intrinsic dispersion of the parent distribution, calculated using the HBM (see text), while the bottom panels show the number of objects in each bin. Both $R_{\rm blur}$ and $R_{\rm pex}$ decrease significantly to higher luminosities, confirming an X-ray Baldwin effect for both the narrow and broad components of the Fe K$\alpha$ line and associated Compton reflection.  Both also show a significant reduction in strength with redshift, and a milder increase with $N_{\rm H}$. These trends might be a by-product of the Baldwin effect.}
\label{fig:BaldwinR}
\end{figure*}

\subsection{Dependence on other parameters}

Our expanded sample compared to that of \citet{Baronchelli2018} gives a more robust detection of the reflection components and thus offers the opportunity to investigate any dependence of the reflection on other parameters. To this end, we computed the fraction of sources with $\mathrm{S/N>7}$ showing $Z_{blur} > Z_{pex}$. This is shown in Figure \ref{fig:Baldwin} as function of luminosity, redshift and $\mathrm{N_H}$ bins. Errors are calculated assuming that the fractions follow a binomial distribution an the bin size is chosen to have the same interval size for the parameter on the x-axis.

In Fig.~\ref{fig:Baldwin}, we see that the fraction of sources showing larger evidence for broad reflection decreases with increasing luminosity (see Figure \ref{fig:Baldwin}, left panel). In the plot, we remove the 12 sources with luminosity less than $\mathrm{L<10^{41}}$ erg/s. The excluded sources are the 12 with the lowest luminosity in the sample. They cover a 3 dex luminosity interval and hence the lowest luminosity bin in the Baldwin plot would be severely underpopulated with large error bars. In addition there is a minor concern that at these very low luminosities, sources may be partially contaminated by non-AGN emission. While this is likely to have a negligible effect on the overall sample properties, the reflection properties determined in this lowest luminosity bin may be somewhat unreliable. Moreover, including those sources does not change the conclusions from Figure \ref{fig:Baldwin}.
We also notice an anti-correlation with increasing redshift and a clear increase of the fraction of broadened sources with increasing $\mathrm{N_H}$ (see Figure \ref{fig:Baldwin}, middle and right panel). 

The anti-correlation of the equivalent width (EW) of the Fe K$\mathrm{\alpha}$ line and the luminosity of AGN is well-known characteristic \citep{Iwasawa1993, Nandra1997, Page2003}. This anti-correlation, called the Iwasawa-Taniguchi or X-ray Baldwin effect (Baldwin effect hereafter), has been seen both for the narrow core of the Fe K$\alpha$, and has also been claimed for the broad component of the line \citep{Nandra1997}. The anti-correlation between luminosity and fraction of sources selected as broad (Figure \ref{fig:Baldwin}, left panel) could be a consequence of the Baldwin effect for the broadened component of the Fe K$\mathrm{\alpha}$ line.

To study this phenomenon in more detail, we explore the relationship between the reflection fraction R (see Table \ref{blur}) of both the broad and narrow reflection components in the most complex model and the luminosity of the sample sources (see Figure \ref{fig:BaldwinR}, left). The mean $R$ values and their intrinsic dispersion were calculated using the HBM, as for the mean values for the whole sample, and exclude from the analysis the 12 sources with $\mathrm{L<10^{41}}$ erg/s to avoid contamination from star forming galaxies.

In Figure \ref{fig:BaldwinR}, we see a clear effect that both the narrow and broad Compton reflection fractions decrease significantly as a function of luminosity, from $R\sim 0.5$ at the lowest luminosities to $R\sim 0.1$ at the highest. Thus we confirm the existence of the Baldwin effect for both the broad and narrow components of the line.

Fig.~\ref{fig:BaldwinR} also shows the dependence of the reflection strengths with redshift and $\mathrm{N_H}$. The same trends shown in Figure \ref{fig:Baldwin} are seen, with a reduction in the $R$ values with redshift, and a weak increase seen with obscuration. Both of these trends might be wholly or partially a consequence of the Baldwin effect, given the usual correlation between luminosity and redshift seen in flux-limited samples (Fig.~\ref{fig:Lum}, and the anti-correlation seen between luminosity and obscured fraction \citep{Steffen2003,Barger2005,Ricci2017}.

\begin{center}
   \begin{figure*}
   
     \subfloat{%
        \includegraphics[scale=0.38]{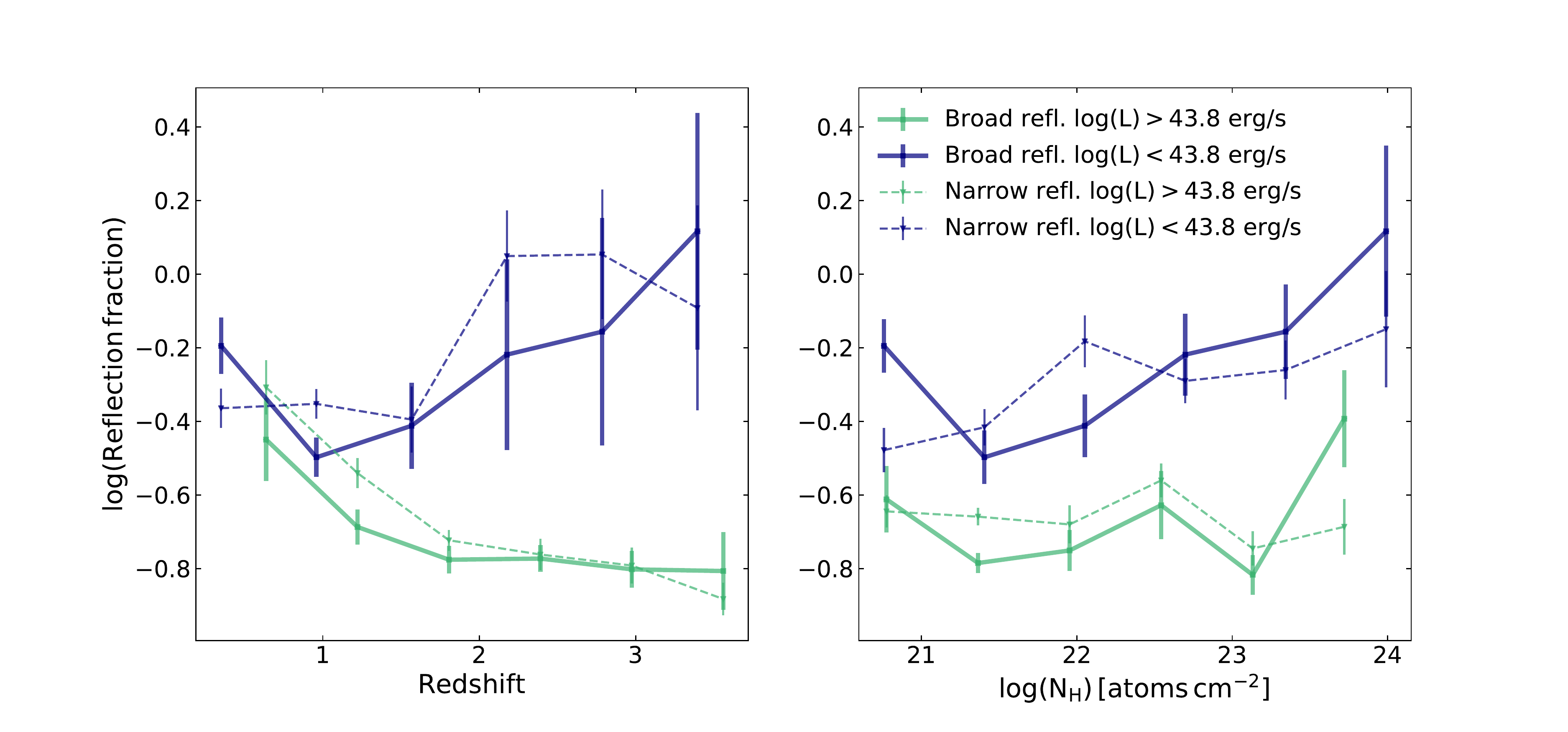}
    }

     \subfloat{%
        \includegraphics[scale=0.38]{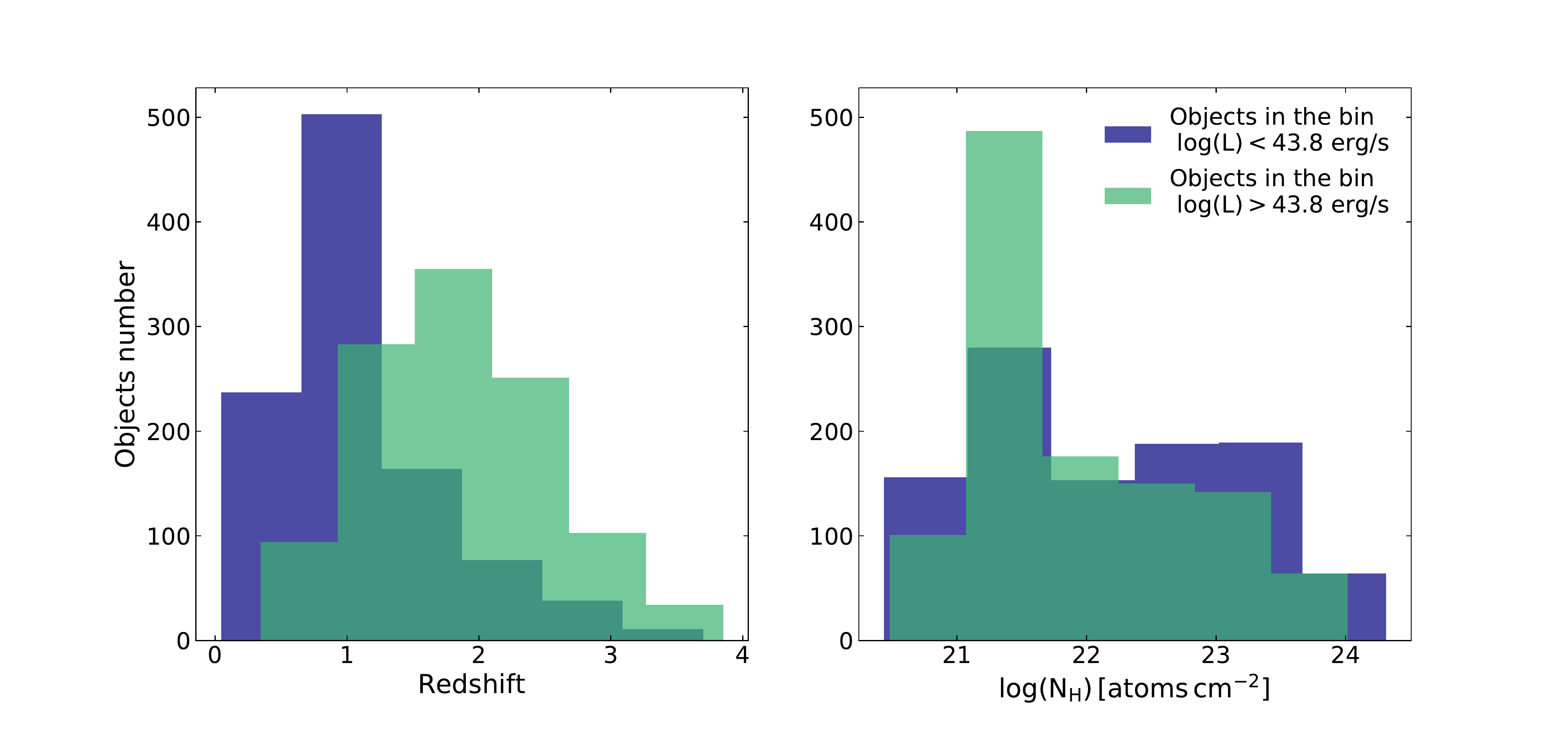}
     }
   \caption{R value of broad (solid line) and narrow (dashed line) reflection as function of redshift (\textit{left}) and column density $\mathrm{N_H}$ (\textit{right}) in for the subsample with $\mathrm{log(L)>43.8 \, erg/s}$ (red) and $\mathrm{log(L)<43.8 \, erg/s}$ (blue). 
   Both broad and narrow reflection component show a smaller R value at higher luminosity. The odd behavior of the R value in the last bin of the low luminosity sub-sample might be induced by low statistic effects, in fact that bin only comprises 11 objects.}
   \label{fig:Lsplit}
   \end{figure*}
\end{center}

We test this hypothesis by splitting the sample (see Figure \ref{fig:Lsplit}) into low-luminosity ($\mathrm{L < 10^{43.8}}$ erg/s, blue in Figure \ref{fig:Lsplit}) and high-luminosity ($\mathrm{L > 10^{43.8}}$ erg/s, red in Figure \ref{fig:Lsplit}) sub-samples. We chose this luminosity threshold because it splits the sample almost in half, with 1042 sources in the low-luminosity sample and 1120 in the high-luminosity one. The solid curves show the behaviour of the reflection fraction of the broad disk reflection component while the dashed lines show the R from the narrow torus reflection. We notice that the curves of R as function of redshift and $\mathrm{N_H}$ seem to flatten for higher luminosities, thus the trends of R vs. redshift and $\mathrm{N_H}$ might be indeed be mirroring the dependency of R with luminosity. Even in Figure \ref{fig:Lsplit}, the stronger trend we can observe is the one with luminosity. In fact, the R values of broad and narrow reflection components at luminosities of $\mathrm{L > 10^{43.8}}$ erg/s are consistently lower than the R values at $\mathrm{L < 10^{43.8}}$ erg/s, as the X-ray Baldwin effect would predict.

\section{Discussion}
\label{discussion}

In this work we have used the deepest X-ray fields performed by \textit{Chandra} to place constraints on 
X-ray Compton reflection in a sample of AGN typical of the overall population, covering a luminosity range $\log L=41-45$ erg/s out to $z\sim 4$. We build on the previous work by \citet{Baronchelli2018}, adopting the same Bayesian framework BXA to fit the spectra, determine parameters, and compare models. We confirm strong evidence for Compton reflection, and by implication also iron K$\alpha$ emission, both from distant material, most likely the torus envisaged in orientation-dependent unification schemes, and relativistically broadened reflection modelled as arising from the inner accretion disk. 

This relativistic reflection is expected to be ubiquitous in the standard scenario where the SMBH is surrounded by an accretion disk and a hot corona of electrons. As such, perhaps the most important result of our study is to confirm that paradigm in the general population of AGN, which are responsible for the bulk of black hole growth in the Universe, and the majority of the X-ray background radiation \citep[e.g.][]{Aird2015,Buchner2015}. We observe that the model 
with maximally spinning Kerr BH is preferred over a model with Schwarzschild BH, reproducing the result from \citet{Baronchelli2018}. This confirms that a portion of the reflection comes from very close to the black hole, indeed perhaps from within the $6 R_{\rm g}$ innermost stable circular orbit of a non-rotating black hole. In turn this implies both that a relatively cool accretion disk extends within this radius, and that the X-ray emission comes from the innermost regions, and is compact enough that a substantial proportion of the disk illumination is at these small radii.
The result is supported by studies of microlensed quasars at high redshift. For example, \citet{Dai2019} show evidence of high spin ($\mathrm{a>0.8}$) in a sample of five lensed quasars at high redshift ($\mathrm{z > 1.2}$) and an ultra-compact X-ray emitting region with size $<10 R_{\rm g}$.

This provides one possible explanation for one of the more puzzling aspects of our analysis, that being the average strength of the relativistic reflection component. We find a value for the average reflection fraction of $R_{\rm blur} \sim 0.3$, contrasting with the $R_{\rm blur} \sim 1$ expected for a flat disk illuminated by a point X-ray source. This relatively weak blurred reflection, at least on average, is in agreement with observations of local AGN \citep{Nandra2007}, although some nearby AGN also show anomalously strong reflection \citep[e.g. MCG-6-30-16 and NGC 1365][]{Fabian2002,Risaliti2013}.  
Both of these facts can be explained by strong relativistic effects and in particular light bending close to the central black hole. This can result in the reflection being either weaker, or stronger than that expected for a flat disc depending on geometrical considerations \citep{Miniutti2004}. A reduced reflection strength would also be expected if the X-ray emission is beamed away from the disk \citep{Beloborodov1999}. 

Apparently weak reflection may also be explained by ionization of the disk. Some reduction of the line flux is expected for moderate ionization due to resonant trapping, and at very high ionization parameters no line is produced at all once iron becomes fully ionized \citep{Ross1993}. In this case the reflection continuum would also become hard to distinguish from the primary continuum. 
 
There may also be geometric effects. The strength of the reflection component is maximised for our assumed geometry of a semi-infinite slab illuminated by a point source. If the real accretion disk-corona geometry is different to this then the reflection is reduced. For example, if the corona of hot electrons were at a height $h$ comparable than the ISCO radius $r_{ms}$, fewer photons from the corona would intercept the disk. The divergence from the "lamp post" supported is corroborated by studies of microlensed quasars \citep[e.g.][]{Chartas2012,Dai2019}, and furthermore suggest a very compact primary X-ray source, consistent with this idea. This effect would be further exacerbated if the accretion disk is truncated before the last stable orbit, although in this case the relativistic signatures would be less prominent, contrary to the strong evidence for their presence found in this work. 

Our expanded sample compared to that of \citet{Baronchelli2018} has enabled an analysis of the dependence of the strength of reflection from both the torus and the accretion disk with luminosity (see Figure \ref{fig:BaldwinR}, left). We confirm an X-ray "Baldwin effect" in which the strength of the reflection component, and by implication the equivalent with of the iron K$\alpha$ line, reduces with luminosity. 

This effect is quite well established for the narrow core of the iron K$\alpha$ line. The most common interpretation is that of the "receding torus" in which the covering fraction reduces with luminosity (e.g. Simpson 2005). This is in agreement with the observation of a higher fraction of optical type 1 galaxies, and lower prevalence of X-ray absorption, at high luminosities \citep[e.g.][]{Ueda2003,Hasinger2005,Buchner2015}. 

The possible presence of a Baldwin effect for the broad part of the emission line was suggested by \citet{Nandra1997b}. This should not have the same physical origin as the narrow-line X-ray Baldwin effect if the broad line comes from the accretion disk, rather than the torus. The near-absence of reflection in the highest luminosity objects may in part explain why the average reflection fraction in our sample is so low, and hence might be due to the same effects e.g. geometry, special or general relativistic beaming and/or disk ionization. A priori it is difficult to see where there should be a strong relationship between the disk-corona geometry and the luminosity, disfavouring this interpretation. On the other hand, if photons are beamed away from the disk this would result in an enhanced luminosity for a given object, along with weaker reflection, as observed. At higher luminosities, the disk may also be more highly ionized suppressing the iron K$\alpha$ line and reflection continuum, as discussed above, and providing a natural explanation for the Baldwin effect. 

Looking next to the apparent dependence of $R$ with redshift, this seems most likely to be a consequence of the Baldwin effect, given the very strong luminosity-redshift correlation in our flux-limited samples. We therefore do not speculate further on the possibility of evolution of the disk-corona system over cosmic time, but if this can be confirmed with better data and samples it would certainly be an intriguing phenomenon. 

We observe also that the fraction of sources selected as broad and the R value of both disk and torus reflection shows a mild increase with $\mathrm{N_H}$ (see Figures \ref{fig:Baldwin} and \ref{fig:BaldwinR}, right panel).
At face value this is the opposite of what might be expected, because in standard orientation-dependent unification schemes, absorbed sources should be seen at high inclination, whence the observed reflection signatures are weaker. The observed increase could also partially be a consequence of the Baldwin effect. As discussed above lower luminosity AGN are more likely to be absorbed, and when we restrict the luminosity range in our analysis the effect does indeed seem weaker. 

A further effect is the possible degeneracy of the broad iron line with complex absortion in sources with high $\mathrm{N_H}$. If the absorption is in fact more complicate than the simple model assumed here, then for moderately high values around $\mathrm{\log N_H}=23$, mismodelling could introduce apparent curvature in the continuum in the 5-6 keV range, mimicking a broad red wing to the line. Unfortunately, the quality of the spectra in our sample prevents us from constraining the disk inclination parameter for most of the sources. Thus we can neither confirm nor rule out the degeneracy of the obscuration with the inclination of the disk on this basis.

One caveat to the above discussion is that is clearly challenging with spectra of the quality used in this work to decompose the reflection into its broad and narrow components. It can be noted from Figure \ref{fig:BaldwinR} (left), for example, that the luminosity dependence of the $R$ value for the broad disk reflection and for the narrow reflection from the torus have a very similar behavior. Since the two components arise from very different regions around the SMBH, this may indicates that the component are not well decoupled in the model, and that there is considerable co-variance between them.

Properly decoupling and measuring the properties of X-ray reflection in individual AGN at high redshift requires an X-ray telescope with significantly higher throughput then the current generation of instrument. Once launched, \textit{Athena} \citep{Nandra2013} will provide this capability. Deep field osbervations with the Athena Wide Field Imager \citep{Rau2013,Meidinger2016} will yield one or two orders of magnitude more photons per unit exposure than \textit{Chandra},
giving high quality spectra for individual objects like those in our sample. Follow-up observations of selected objects with the \textit{Athena} X-ray Integral Field Unit \citep[X-IFU; ][]{Barret2016} of brighter examples found with the WFI will enable the first high resolution spectra of such objects. 

On the other hand, with this project we have confirmed the potential of of X-ray spectroscopy combined with Bayesian inference to reveal information about the population properties of AGN, even with individual spectra of low signal-to-noise ratio. There are some limitations, in that we have also demonstrated that adding spectra with very low signal-to-noise ratio at some points fails to add further information. Once appropriate signal-to-noise ratio cuts are applied, however, each new objects added to the analysis is able to strengthen our inferences about the accretion processes in SMBHs and the gas structures around them. The recent launch of the instrument \textit{eROSITA} \citep{Predehl2010} aboard the SRG satellite present a particularly exciting opportunity to take this forward. eROSITA will detect millions of AGN \citep{Merloni2012, Kolodzig2013} spread over the full sky and filling out an extremely broad luminosity-redshift plane. This will open up new and exciting possibilities for the application of our methods.

\section{Summary and Conclusions}
\label{sumconc}
We present an analysis of the X-ray spectra of sources taken from the four \textit{Chandra} deep fields CDFS 7Ms, CDFN, AEGIS and COSMOS to determine the Compton reflection and iron K$\alpha$ line properties of typical AGN outside the nearby Universe. To this purpose, we fit all the spectra individually using BXA, rather than stacking them. We fit four models of increasing complexity, starting from a simple absorbed power-law and adding to this model a narrow reflection component (\texttt{pexmon}) and subsequently a further relativistically blurred disk reflection (\texttt{kerrconv(pexmon)}) with dimensionless spin parameterfixed at two values, $\mathrm{a=0}$ and $\mathrm{a=0.998}$. We perform simulations to determine from which signal-to-noise ratio, S/N, a source adds information to the total Bayesian evidence of the sample, finding that below a S/N of 7 the sources are too faint to add any new information, so we restrict the bulk of our analysis to sources above this limit. The outputs of BXA are the best fit parameters for and the Bayesian evidence of the model, thus we can use the latter to calculate the Bayes factor (BF) for model comparison. Based on the fits to the individual spectra, we adopt a hierarchical Bayesian model (see Appendix \ref{HBM}) to determine the sample properties. 

\vspace{0.5cm}
Our main findings are:
\begin{itemize}
    \item When considering the sample as a whole, the Bayesian evidence comparison shows a preference for a model containing both narrow and broad, relativistic Compton reflection. This is in agreement with the result of \citet{Baronchelli2018} using a smaller sample of spectra from the CDFS 4Ms.
    \item As in \citet{Baronchelli2018}, we find that the broad disk reflection model with a maximally spinning BH is preferred over one with spin=0.   
    \item The HBM shows that on average both narrow and broad Compton reflection is relatively weak. We find a mean of $\mathrm{log(R_{pex}) = -0.53}$ with spread $\mathrm{\sigma_{log(R_{pex})} = 0.2}$ for the narrow reflection component and mean of $\mathrm{log(R_{blur}) = -0.57}$ with spread of $\mathrm{\sigma_{log(R_{blur})} = 0.14}$ for the blurred reflection component. This implies a departure from the simple "lamp post" geometry assumed in our reflection models, and in the case of the broad reflection possibly light bending or ionization effects. 
    \item We investigate the presence of an X-ray Baldwin effect in our sample, confirming a decrease in the reflection strength for both the distant and blurred components, and by inference both the narrow and broad components of the iron K$\alpha$ line. The former may be explained by a "receding torus" model, whereas as the latter implies a dependence of the inner disk geometry and or ionization with luminosity. 
    \item We also find anti-correlations of the reflection fraction of the disk and torus with redshift, and a weak positive correlation with $\mathrm{N_H}$. Both may, however, be artifacts of the Baldwin effect. 
    \item With this analysis, we confirm the power of Bayesian statistics to infer important physical characteristics and features of AGN using a sample of relatively low S/N X-ray spectra, a technique which can be applied powerfully to the upcoming eROSITA survey. To measure the properties of X-ray reflection in individual high-redshift AGN we will need instrument with significantly higher effective area, such \textit{Athena}.
\end{itemize}

\section*{Acknowledgements}

We thank the referee, Francisco Carrerra, for a number of helpful suggestions which considerably improved this manuscript.

This research made use of {\tt Astropy}, a community-developed core Python package for Astronomy \citep{Astropy2013} and the NASA's Astrophysics Data System. This research made use of APLpy, an open-source plotting package for Python \citep{Robitaille2012}. We also used extensively the Python package Matplotlib \citep{Hunter2007}.

JB acknowledges support from the CONICYT-Chile grants Basal-CATA PFB-06/2007 $\&$ AFB-170002, FONDECYT Postdoctorados 3160439 and the Ministry of Economy, Development, and Tourism's Millennium Science Initiative through grant IC120009, awarded to The Millennium Institute of Astrophysics, MAS. 
This research was supported by the DFG cluster of excellence ,,Origin and Structure of the Universe''.

\section*{Data availability}
The data underlying this article will be shared on a reasonable request to the corresponding author. Part of the data (the posterior distribution of log(R)) can be found here: \url{https://github.com/blinda/HierarchicalBayes}.



\bibliographystyle{mnras}
\bibliography{megaSample} 

\begin{thebibliography}{}
\makeatletter
\relax
\def\mn@urlcharsother{\let\do\@makeother \do\$\do\&\do\#\do\^\do\_\do\%\do\~}
\def\mn@doi{\begingroup\mn@urlcharsother \@ifnextchar [ {\mn@doi@}
  {\mn@doi@[]}}
\def\mn@doi@[#1]#2{\def\@tempa{#1}\ifx\@tempa\@empty \href
  {http://dx.doi.org/#2} {doi:#2}\else \href {http://dx.doi.org/#2} {#1}\fi
  \endgroup}
\def\mn@eprint#1#2{\mn@eprint@#1:#2::\@nil}
\def\mn@eprint@arXiv#1{\href {http://arxiv.org/abs/#1} {{\tt arXiv:#1}}}
\def\mn@eprint@dblp#1{\href {http://dblp.uni-trier.de/rec/bibtex/#1.xml}
  {dblp:#1}}
\def\mn@eprint@#1:#2:#3:#4\@nil{\def\@tempa {#1}\def\@tempb {#2}\def\@tempc
  {#3}\ifx \@tempc \@empty \let \@tempc \@tempb \let \@tempb \@tempa \fi \ifx
  \@tempb \@empty \def\@tempb {arXiv}\fi \@ifundefined
  {mn@eprint@\@tempb}{\@tempb:\@tempc}{\expandafter \expandafter \csname
  mn@eprint@\@tempb\endcsname \expandafter{\@tempc}}}

\bibitem[\protect\citeauthoryear{{Aird} et~al.,}{{Aird}
  et~al.}{2010}]{Aird2010}
{Aird} J.,  et~al., 2010, \mn@doi [\mnras] {10.1111/j.1365-2966.2009.15829.x},
  \href {https://ui.adsabs.harvard.edu/abs/2010MNRAS.401.2531A} {401, 2531}

\bibitem[\protect\citeauthoryear{{Aird}, {Coil}, {Georgakakis}, {Nandra},
  {Barro}  \& {P{\'e}rez-Gonz{\'a}lez}}{{Aird} et~al.}{2015}]{Aird2015}
{Aird} J.,  {Coil} A.~L.,  {Georgakakis} A.,  {Nandra} K.,  {Barro} G.,
  {P{\'e}rez-Gonz{\'a}lez} P.~G.,  2015, \mn@doi [\mnras]
  {10.1093/mnras/stv1062}, \href
  {https://ui.adsabs.harvard.edu/abs/2015MNRAS.451.1892A} {451, 1892}

\bibitem[\protect\citeauthoryear{{Arnaud}}{{Arnaud}}{1996}]{Arnaud1996}
{Arnaud} K.~A.,  1996, in {Jacoby} G.~H.,  {Barnes} J.,  eds,  Astronomical
  Society of the Pacific Conference Series Vol. 101, Astronomical Data Analysis
  Software and Systems V. p.~17

\bibitem[\protect\citeauthoryear{{Astropy Collaboration} et~al.,}{{Astropy
  Collaboration} et~al.}{2013}]{Astropy2013}
{Astropy Collaboration} et~al., 2013, \mn@doi [\aap]
  {10.1051/0004-6361/201322068}, \href
  {http://adsabs.harvard.edu/abs/2013A%26A...558A..33A} {558, A33}

\bibitem[\protect\citeauthoryear{{Barger}, {Cowie}, {Mushotzky}, {Yang},
  {Wang}, {Steffen}  \& {Capak}}{{Barger} et~al.}{2005}]{Barger2005}
{Barger} A.~J.,  {Cowie} L.~L.,  {Mushotzky} R.~F.,  {Yang} Y.,  {Wang} W.~H.,
  {Steffen} A.~T.,   {Capak} P.,  2005, \mn@doi [\aj] {10.1086/426915}, \href
  {https://ui.adsabs.harvard.edu/abs/2005AJ....129..578B} {129, 578}

\bibitem[\protect\citeauthoryear{{Baronchelli}, {Nandra}  \&
  {Buchner}}{{Baronchelli} et~al.}{2018}]{Baronchelli2018}
{Baronchelli} L.,  {Nandra} K.,   {Buchner} J.,  2018, \mn@doi [\mnras]
  {10.1093/mnras/sty2026}, \href
  {http://adsabs.harvard.edu/abs/2018MNRAS.480.2377B} {480, 2377}

\bibitem[\protect\citeauthoryear{{Barret} et~al.,}{{Barret}
  et~al.}{2016}]{Barret2016}
{Barret} D.,  et~al., 2016, in Society of Photo-Optical Instrumentation
  Engineers (SPIE) Conference Series, Vol.~9905, \procspie.
p. 99052F, \mn@doi{10.1117/12.2232432}

\bibitem[\protect\citeauthoryear{{Beloborodov}}{{Beloborodov}}{1999}]{Beloborodov1999}
{Beloborodov} A.~M.,  1999, \mn@doi [\apjl] {10.1086/311810}, \href
  {https://ui.adsabs.harvard.edu/abs/1999ApJ...510L.123B} {510, L123}

\bibitem[\protect\citeauthoryear{{Betancourt}}{{Betancourt}}{2015}]{Betancourt2015}
{Betancourt} M.~J.,  2015, arXiv e-prints, \href
  {https://ui.adsabs.harvard.edu/abs/2015arXiv150201510B} {p. arXiv:1502.01510}

\bibitem[\protect\citeauthoryear{{Blackburn}}{{Blackburn}}{1995}]{Blackburn1995}
{Blackburn} J.~K.,  1995, in {Shaw} R.~A.,  {Payne} H.~E.,   {Hayes} J. J.~E.,
  eds, Astronomical Society of the Pacific Conference Series, Vol.~77,
  Astronomical Data Analysis Software and Systems IV.
p.~367

\bibitem[\protect\citeauthoryear{{Brenneman} \& {Reynolds}}{{Brenneman} \&
  {Reynolds}}{2006}]{Brenneman2006}
{Brenneman} L.~W.,  {Reynolds} C.~S.,  2006, \mn@doi [\apj] {10.1086/508146},
  \href {http://adsabs.harvard.edu/abs/2006ApJ...652.1028B} {652, 1028}

\bibitem[\protect\citeauthoryear{{Broos}, {Townsley}, {Feigelson}, {Getman},
  {Bauer}  \& {Garmire}}{{Broos} et~al.}{2010}]{Broos2010}
{Broos} P.~S.,  {Townsley} L.~K.,  {Feigelson} E.~D.,  {Getman} K.~V.,  {Bauer}
  F.~E.,   {Garmire} G.~P.,  2010, \mn@doi [\apj]
  {10.1088/0004-637X/714/2/1582}, \href
  {https://ui.adsabs.harvard.edu/abs/2010ApJ...714.1582B} {714, 1582}

\bibitem[\protect\citeauthoryear{{Broos}, {Townsley}, {Getman}  \&
  {Bauer}}{{Broos} et~al.}{2012}]{Broos2012}
{Broos} P.,  {Townsley} L.,  {Getman} K.,   {Bauer} F.,  2012, {AE: ACIS
  Extract} (\mn@eprint {ascl} {1203.001})

\bibitem[\protect\citeauthoryear{{Brusa}, {Gilli}  \& {Comastri}}{{Brusa}
  et~al.}{2005}]{Brusa2005}
{Brusa} M.,  {Gilli} R.,   {Comastri} A.,  2005, \mn@doi [\apjl]
  {10.1086/428928}, \href {http://adsabs.harvard.edu/abs/2005ApJ...621L...5B}
  {621, L5}

\bibitem[\protect\citeauthoryear{Buchner}{Buchner}{2019}]{Buchner2019}
Buchner J.,  2019, \mn@doi [Publications of the Astronomical Society of the
  Pacific] {10.1088/1538-3873/aae7fc}, 131, 108005

\bibitem[\protect\citeauthoryear{{Buchner} et~al.,}{{Buchner}
  et~al.}{2014}]{Buchner2014}
{Buchner} J.,  et~al., 2014, \mn@doi [\aap] {10.1051/0004-6361/201322971},
  \href {https://ui.adsabs.harvard.edu/abs/2014A&A...564A.125B} {564, A125}

\bibitem[\protect\citeauthoryear{{Buchner} et~al.,}{{Buchner}
  et~al.}{2015}]{Buchner2015}
{Buchner} J.,  et~al., 2015, \mn@doi [\apj] {10.1088/0004-637X/802/2/89}, \href
  {http://adsabs.harvard.edu/abs/2015ApJ...802...89B} {802, 89}

\bibitem[\protect\citeauthoryear{{Chartas}, {Kochanek}, {Dai}, {Moore},
  {Mosquera}  \& {Blackburne}}{{Chartas} et~al.}{2012}]{Chartas2012}
{Chartas} G.,  {Kochanek} C.~S.,  {Dai} X.,  {Moore} D.,  {Mosquera} A.~M.,
  {Blackburne} J.~A.,  2012, \mn@doi [\apj] {10.1088/0004-637X/757/2/137},
  \href {https://ui.adsabs.harvard.edu/abs/2012ApJ...757..137C} {757, 137}

\bibitem[\protect\citeauthoryear{{Chaudhary}, {Brusa}, {Hasinger}, {Merloni},
  {Comastri}  \& {Nandra}}{{Chaudhary} et~al.}{2012}]{Chaudhary2012}
{Chaudhary} P.,  {Brusa} M.,  {Hasinger} G.,  {Merloni} A.,  {Comastri} A.,
  {Nandra} K.,  2012, \mn@doi [\aap] {10.1051/0004-6361/201117126}, \href
  {https://ui.adsabs.harvard.edu/abs/2012A&A...537A...6C} {537, A6}

\bibitem[\protect\citeauthoryear{{Civano} et~al.,}{{Civano}
  et~al.}{2016}]{Civano2016}
{Civano} F.,  et~al., 2016, \mn@doi [\apj] {10.3847/0004-637X/819/1/62}, \href
  {https://ui.adsabs.harvard.edu/abs/2016ApJ...819...62C} {819, 62}

\bibitem[\protect\citeauthoryear{{Comastri}, {Brusa}  \& {Civano}}{{Comastri}
  et~al.}{2004}]{Comastri2004}
{Comastri} A.,  {Brusa} M.,   {Civano} F.,  2004, \mn@doi [\mnras]
  {10.1111/j.1365-2966.2004.07929.x}, \href
  {http://adsabs.harvard.edu/abs/2004MNRAS.351L...9C} {351, L9}

\bibitem[\protect\citeauthoryear{{Corral} et~al.,}{{Corral}
  et~al.}{2008}]{Corral2008}
{Corral} A.,  et~al., 2008, \mn@doi [\aap] {10.1051/0004-6361:200810168}, \href
  {http://adsabs.harvard.edu/abs/2008A%26A...492...71C} {492, 71}

\bibitem[\protect\citeauthoryear{{Dai}, {Steele}, {Guerras}, {Morgan}  \&
  {Chen}}{{Dai} et~al.}{2019}]{Dai2019}
{Dai} X.,  {Steele} S.,  {Guerras} E.,  {Morgan} C.~W.,   {Chen} B.,  2019,
  \mn@doi [\apj] {10.3847/1538-4357/ab1d56}, \href
  {https://ui.adsabs.harvard.edu/abs/2019ApJ...879...35D} {879, 35}

\bibitem[\protect\citeauthoryear{{Elvis} et~al.,}{{Elvis}
  et~al.}{1994}]{Elvis1994}
{Elvis} M.,  et~al., 1994, \mn@doi [\apjs] {10.1086/192093}, \href
  {https://ui.adsabs.harvard.edu/abs/1994ApJS...95....1E} {95, 1}

\bibitem[\protect\citeauthoryear{{Fabian}, {Rees}, {Stella}  \&
  {White}}{{Fabian} et~al.}{1989}]{Fabian1989}
{Fabian} A.~C.,  {Rees} M.~J.,  {Stella} L.,   {White} N.~E.,  1989, \mn@doi
  [\mnras] {10.1093/mnras/238.3.729}, \href
  {https://ui.adsabs.harvard.edu/abs/1989MNRAS.238..729F} {238, 729}

\bibitem[\protect\citeauthoryear{{Fabian}, {Iwasawa}, {Reynolds}  \&
  {Young}}{{Fabian} et~al.}{2000}]{Fabian2000}
{Fabian} A.~C.,  {Iwasawa} K.,  {Reynolds} C.~S.,   {Young} A.~J.,  2000,
  \mn@doi [\pasp] {10.1086/316610}, \href
  {http://adsabs.harvard.edu/abs/2000PASP..112.1145F} {112, 1145}

\bibitem[\protect\citeauthoryear{{Fabian} et~al.,}{{Fabian}
  et~al.}{2002}]{Fabian2002}
{Fabian} A.~C.,  et~al., 2002, \mn@doi [\mnras]
  {10.1046/j.1365-8711.2002.05740.x}, \href
  {https://ui.adsabs.harvard.edu/abs/2002MNRAS.335L...1F} {335, L1}

\bibitem[\protect\citeauthoryear{{Falocco}, {Carrera}, {Corral}, {Laird},
  {Nandra}, {Barcons}, {Page}  \& {Digby-North}}{{Falocco}
  et~al.}{2012}]{Falocco2012}
{Falocco} S.,  {Carrera} F.~J.,  {Corral} A.,  {Laird} E.,  {Nandra} K.,
  {Barcons} X.,  {Page} M.~J.,   {Digby-North} J.,  2012, \mn@doi [\aap]
  {10.1051/0004-6361/201117965}, \href
  {https://ui.adsabs.harvard.edu/abs/2012A&A...538A..83F} {538, A83}

\bibitem[\protect\citeauthoryear{{Falocco} et~al.,}{{Falocco}
  et~al.}{2013}]{Falocco2013}
{Falocco} S.,  et~al., 2013, \mn@doi [\aap] {10.1051/0004-6361/201321083},
  \href {http://adsabs.harvard.edu/abs/2013A%26A...555A..79F} {555, A79}

\bibitem[\protect\citeauthoryear{{Falocco}, {Carrera}, {Barcons}, {Miniutti}
  \& {Corral}}{{Falocco} et~al.}{2014}]{Falocco2014}
{Falocco} S.,  {Carrera} F.~J.,  {Barcons} X.,  {Miniutti} G.,   {Corral} A.,
  2014, \mn@doi [\aap] {10.1051/0004-6361/201322812}, \href
  {http://adsabs.harvard.edu/abs/2014A%26A...568A..15F} {568, A15}

\bibitem[\protect\citeauthoryear{{Feroz} \& {Hobson}}{{Feroz} \&
  {Hobson}}{2008}]{Feroz2008}
{Feroz} F.,  {Hobson} M.~P.,  2008, \mn@doi [\mnras]
  {10.1111/j.1365-2966.2007.12353.x}, \href
  {http://adsabs.harvard.edu/abs/2008MNRAS.384..449F} {384, 449}

\bibitem[\protect\citeauthoryear{{Feroz}, {Hobson}  \& {Bridges}}{{Feroz}
  et~al.}{2009}]{Feroz2009}
{Feroz} F.,  {Hobson} M.~P.,   {Bridges} M.,  2009, \mn@doi [\mnras]
  {10.1111/j.1365-2966.2009.14548.x}, \href
  {http://adsabs.harvard.edu/abs/2009MNRAS.398.1601F} {398, 1601}

\bibitem[\protect\citeauthoryear{{Feroz}, {Hobson}, {Cameron}  \&
  {Pettitt}}{{Feroz} et~al.}{2013}]{Feroz2013}
{Feroz} F.,  {Hobson} M.~P.,  {Cameron} E.,   {Pettitt} A.~N.,  2013, preprint,
  \href {http://adsabs.harvard.edu/abs/2013arXiv1306.2144F} {} (\mn@eprint
  {arXiv} {1306.2144})

\bibitem[\protect\citeauthoryear{{Fruscione} et~al.,}{{Fruscione}
  et~al.}{2006}]{Fruscione2006}
{Fruscione} A.,  et~al., 2006, in Society of Photo-Optical Instrumentation
  Engineers (SPIE) Conference Series, Vol.~6270, \procspie.
p. 62701V, \mn@doi{10.1117/12.671760}

\bibitem[\protect\citeauthoryear{{Georgakakis} et~al.,}{{Georgakakis}
  et~al.}{2011}]{Georgakakis2011}
{Georgakakis} A.,  et~al., 2011, \mn@doi [\mnras]
  {10.1111/j.1365-2966.2011.19650.x}, \href
  {http://adsabs.harvard.edu/abs/2011MNRAS.418.2590G} {418, 2590}

\bibitem[\protect\citeauthoryear{{George} \& {Fabian}}{{George} \&
  {Fabian}}{1991}]{George1991}
{George} I.~M.,  {Fabian} A.~C.,  1991, \mn@doi [\mnras]
  {10.1093/mnras/249.2.352}, \href
  {http://adsabs.harvard.edu/abs/1991MNRAS.249..352G} {249, 352}

\bibitem[\protect\citeauthoryear{{Ghisellini}, {Haardt}  \&
  {Matt}}{{Ghisellini} et~al.}{1994}]{Ghisellini1994}
{Ghisellini} G.,  {Haardt} F.,   {Matt} G.,  1994, \mn@doi [\mnras]
  {10.1093/mnras/267.3.743}, \href
  {https://ui.adsabs.harvard.edu/abs/1994MNRAS.267..743G} {267, 743}

\bibitem[\protect\citeauthoryear{{Haardt} \& {Maraschi}}{{Haardt} \&
  {Maraschi}}{1991}]{Haardt1991}
{Haardt} F.,  {Maraschi} L.,  1991, \mn@doi [\apjl] {10.1086/186171}, \href
  {https://ui.adsabs.harvard.edu/abs/1991ApJ...380L..51H} {380, L51}

\bibitem[\protect\citeauthoryear{{Hasinger}, {Miyaji}  \& {Schmidt}}{{Hasinger}
  et~al.}{2005}]{Hasinger2005}
{Hasinger} G.,  {Miyaji} T.,   {Schmidt} M.,  2005, \mn@doi [\aap]
  {10.1051/0004-6361:20042134}, \href
  {https://ui.adsabs.harvard.edu/abs/2005A&A...441..417H} {441, 417}

\bibitem[\protect\citeauthoryear{{Hsu} et~al.,}{{Hsu} et~al.}{2014}]{Hsu2014}
{Hsu} L.-T.,  et~al., 2014, \mn@doi [\apj] {10.1088/0004-637X/796/1/60}, \href
  {https://ui.adsabs.harvard.edu/abs/2014ApJ...796...60H} {796, 60}

\bibitem[\protect\citeauthoryear{Hunter}{Hunter}{2007}]{Hunter2007}
Hunter J.~D.,  2007, \mn@doi [Computing In Science \& Engineering]
  {10.1109/MCSE.2007.55}, 9, 90

\bibitem[\protect\citeauthoryear{{Iwasawa} \& {Taniguchi}}{{Iwasawa} \&
  {Taniguchi}}{1993}]{Iwasawa1993}
{Iwasawa} K.,  {Taniguchi} Y.,  1993, \mn@doi [\apjl] {10.1086/186948}, \href
  {https://ui.adsabs.harvard.edu/abs/1993ApJ...413L..15I} {413, L15}

\bibitem[\protect\citeauthoryear{{King} \& {Pringle}}{{King} \&
  {Pringle}}{2006}]{King2006}
{King} A.~R.,  {Pringle} J.~E.,  2006, \mn@doi [\mnras]
  {10.1111/j.1745-3933.2006.00249.x}, \href
  {https://ui.adsabs.harvard.edu/abs/2006MNRAS.373L..90K} {373, L90}

\bibitem[\protect\citeauthoryear{{Kolodzig}, {Gilfanov}, {Sunyaev}, {Sazonov}
  \& {Brusa}}{{Kolodzig} et~al.}{2013}]{Kolodzig2013}
{Kolodzig} A.,  {Gilfanov} M.,  {Sunyaev} R.,  {Sazonov} S.,   {Brusa} M.,
  2013, \mn@doi [\aap] {10.1051/0004-6361/201220880}, \href
  {https://ui.adsabs.harvard.edu/abs/2013A&A...558A..89K} {558, A89}

\bibitem[\protect\citeauthoryear{{Komatsu} et~al.,}{{Komatsu}
  et~al.}{2011}]{Komatsu2011}
{Komatsu} E.,  et~al., 2011, \mn@doi [\apjs] {10.1088/0067-0049/192/2/18},
  \href {http://adsabs.harvard.edu/abs/2011ApJS..192...18K} {192, 18}

\bibitem[\protect\citeauthoryear{{Krolik}, {Madau}  \& {Zycki}}{{Krolik}
  et~al.}{1994}]{Krolik1994}
{Krolik} J.~H.,  {Madau} P.,   {Zycki} P.~T.,  1994, \mn@doi [\apjl]
  {10.1086/187162}, \href
  {https://ui.adsabs.harvard.edu/abs/1994ApJ...420L..57K} {420, L57}

\bibitem[\protect\citeauthoryear{{Laor}}{{Laor}}{1991}]{Laor1991}
{Laor} A.,  1991, \mn@doi [\apj] {10.1086/170257}, \href
  {https://ui.adsabs.harvard.edu/abs/1991ApJ...376...90L} {376, 90}

\bibitem[\protect\citeauthoryear{{Laor} \& {Netzer}}{{Laor} \&
  {Netzer}}{1989}]{Laor1989}
{Laor} A.,  {Netzer} H.,  1989, \mn@doi [\mnras] {10.1093/mnras/238.3.897},
  \href {https://ui.adsabs.harvard.edu/abs/1989MNRAS.238..897L} {238, 897}

\bibitem[\protect\citeauthoryear{{Li} \& {Ma}}{{Li} \& {Ma}}{1983}]{Li1983}
{Li} T.~P.,  {Ma} Y.~Q.,  1983, \mn@doi [\apj] {10.1086/161295}, \href
  {https://ui.adsabs.harvard.edu/abs/1983ApJ...272..317L} {272, 317}

\bibitem[\protect\citeauthoryear{{Liu}, {Yuan}, {Lu}, {Carrera}, {Falocco}  \&
  {Dong}}{{Liu} et~al.}{2016}]{Liu2016}
{Liu} Z.,  {Yuan} W.,  {Lu} Y.,  {Carrera} F.~J.,  {Falocco} S.,   {Dong}
  X.-B.,  2016, \mn@doi [\mnras] {10.1093/mnras/stw2042}, \href
  {http://adsabs.harvard.edu/abs/2016MNRAS.463..684L} {463, 684}

\bibitem[\protect\citeauthoryear{{Luo} et~al.,}{{Luo} et~al.}{2017}]{Luo2017}
{Luo} B.,  et~al., 2017, \mn@doi [\apjs] {10.3847/1538-4365/228/1/2}, \href
  {http://adsabs.harvard.edu/abs/2017ApJS..228....2L} {228, 2}

\bibitem[\protect\citeauthoryear{{Magdziarz} \& {Zdziarski}}{{Magdziarz} \&
  {Zdziarski}}{1995}]{Magdziarz1995}
{Magdziarz} P.,  {Zdziarski} A.~A.,  1995, \mn@doi [\mnras]
  {10.1093/mnras/273.3.837}, \href
  {http://adsabs.harvard.edu/abs/1995MNRAS.273..837M} {273, 837}

\bibitem[\protect\citeauthoryear{{Malkan}}{{Malkan}}{1983}]{Malkan1983}
{Malkan} M.~A.,  1983, \mn@doi [\apj] {10.1086/160981}, \href
  {https://ui.adsabs.harvard.edu/abs/1983ApJ...268..582M} {268, 582}

\bibitem[\protect\citeauthoryear{{Mantovani}, {Nandra}  \& {Ponti}}{{Mantovani}
  et~al.}{2016}]{Mantovani2016}
{Mantovani} G.,  {Nandra} K.,   {Ponti} G.,  2016, \mn@doi [\mnras]
  {10.1093/mnras/stw596}, \href
  {https://ui.adsabs.harvard.edu/abs/2016MNRAS.458.4198M} {458, 4198}

\bibitem[\protect\citeauthoryear{{Matt}}{{Matt}}{2002}]{Matt2002}
{Matt} G.,  2002, \mn@doi [\mnras] {10.1046/j.1365-8711.2002.05890.x}, \href
  {http://adsabs.harvard.edu/abs/2002MNRAS.337..147M} {337, 147}

\bibitem[\protect\citeauthoryear{{Meidinger}, {Eder}, {Eraerds}, {Nand ra},
  {Pietschner}, {Plattner}, {Rau}  \& {Strecker}}{{Meidinger}
  et~al.}{2016}]{Meidinger2016}
{Meidinger} N.,  {Eder} J.,  {Eraerds} T.,  {Nand ra} K.,  {Pietschner} D.,
  {Plattner} M.,  {Rau} A.,   {Strecker} R.,  2016, in Society of Photo-Optical
  Instrumentation Engineers (SPIE) Conference Series, Vol.~9905, \procspie.
p. 99052A, \mn@doi{10.1117/12.2231604}

\bibitem[\protect\citeauthoryear{{Merloni} et~al.,}{{Merloni}
  et~al.}{2012}]{Merloni2012}
{Merloni} A.,  et~al., 2012, preprint, \href
  {http://adsabs.harvard.edu/abs/2012arXiv1209.3114M} {} (\mn@eprint {arXiv}
  {1209.3114})

\bibitem[\protect\citeauthoryear{{Miniutti} \& {Fabian}}{{Miniutti} \&
  {Fabian}}{2004}]{Miniutti2004}
{Miniutti} G.,  {Fabian} A.~C.,  2004, \mn@doi [\mnras]
  {10.1111/j.1365-2966.2004.07611.x}, \href
  {https://ui.adsabs.harvard.edu/abs/2004MNRAS.349.1435M} {349, 1435}

\bibitem[\protect\citeauthoryear{{Nandra}}{{Nandra}}{2006}]{Nandra2006}
{Nandra} K.,  2006, \mn@doi [\mnras] {10.1111/j.1745-3933.2006.00158.x}, \href
  {https://ui.adsabs.harvard.edu/abs/2006MNRAS.368L..62N} {368, L62}

\bibitem[\protect\citeauthoryear{{Nandra} \& {Pounds}}{{Nandra} \&
  {Pounds}}{1994}]{Nandra1994}
{Nandra} K.,  {Pounds} K.~A.,  1994, \mn@doi [\mnras]
  {10.1093/mnras/268.2.405}, \href
  {https://ui.adsabs.harvard.edu/abs/1994MNRAS.268..405N} {268, 405}

\bibitem[\protect\citeauthoryear{{Nandra}, {George}, {Mushotzky}, {Turner}  \&
  {Yaqoob}}{{Nandra} et~al.}{1997a}]{Nandra1997}
{Nandra} K.,  {George} I.~M.,  {Mushotzky} R.~F.,  {Turner} T.~J.,   {Yaqoob}
  T.,  1997a, \mn@doi [\apj] {10.1086/303721}, \href
  {http://adsabs.harvard.edu/abs/1997ApJ...477..602N} {477, 602}

\bibitem[\protect\citeauthoryear{{Nandra}, {George}, {Mushotzky}, {Turner}  \&
  {Yaqoob}}{{Nandra} et~al.}{1997b}]{Nandra1997b}
{Nandra} K.,  {George} I.~M.,  {Mushotzky} R.~F.,  {Turner} T.~J.,   {Yaqoob}
  T.,  1997b, \mn@doi [\apjl] {10.1086/310937}, \href
  {https://ui.adsabs.harvard.edu/abs/1997ApJ...488L..91N} {488, L91}

\bibitem[\protect\citeauthoryear{{Nandra}, {George}, {Mushotzky}, {Turner}  \&
  {Yaqoob}}{{Nandra} et~al.}{1999}]{Nandra1999}
{Nandra} K.,  {George} I.~M.,  {Mushotzky} R.~F.,  {Turner} T.~J.,   {Yaqoob}
  T.,  1999, \mn@doi [\apjl] {10.1086/312252}, \href
  {http://adsabs.harvard.edu/abs/1999ApJ...523L..17N} {523, L17}

\bibitem[\protect\citeauthoryear{{Nandra}, {O'Neill}, {George}  \&
  {Reeves}}{{Nandra} et~al.}{2007}]{Nandra2007}
{Nandra} K.,  {O'Neill} P.~M.,  {George} I.~M.,   {Reeves} J.~N.,  2007,
  \mn@doi [\mnras] {10.1111/j.1365-2966.2007.12331.x}, \href
  {http://adsabs.harvard.edu/abs/2007MNRAS.382..194N} {382, 194}

\bibitem[\protect\citeauthoryear{{Nandra} et~al.,}{{Nandra}
  et~al.}{2013}]{Nandra2013}
{Nandra} K.,  et~al., 2013, preprint, \href
  {http://adsabs.harvard.edu/abs/2013arXiv1306.2307N} {} (\mn@eprint {arXiv}
  {1306.2307})

\bibitem[\protect\citeauthoryear{{Nandra} et~al.,}{{Nandra}
  et~al.}{2015}]{Nandra2015}
{Nandra} K.,  et~al., 2015, \mn@doi [\apjs] {10.1088/0067-0049/220/1/10}, \href
  {http://adsabs.harvard.edu/abs/2015ApJS..220...10N} {220, 10}

\bibitem[\protect\citeauthoryear{Nesseris \& Garc{\'{\i}}a-Bellido}{Nesseris \&
  Garc{\'{\i}}a-Bellido}{2013}]{Nesseris2013}
Nesseris S.,  Garc{\'{\i}}a-Bellido J.,  2013, \mn@doi [Journal of Cosmology
  and Astroparticle Physics] {10.1088/1475-7516/2013/08/036}, 2013, 036

\bibitem[\protect\citeauthoryear{{Page}, {Davis}  \& {Salvi}}{{Page}
  et~al.}{2003}]{Page2003}
{Page} M.~J.,  {Davis} S.~W.,   {Salvi} N.~J.,  2003, \mn@doi [\mnras]
  {10.1046/j.1365-8711.2003.06756.x}, \href
  {https://ui.adsabs.harvard.edu/abs/2003MNRAS.343.1241P} {343, 1241}

\bibitem[\protect\citeauthoryear{{Predehl} et~al.,}{{Predehl}
  et~al.}{2010}]{Predehl2010}
{Predehl} P.,  et~al., 2010, in Space Telescopes and Instrumentation 2010:
  Ultraviolet to Gamma Ray. p. 77320U (\mn@eprint {arXiv} {1001.2502}),
  \mn@doi{10.1117/12.856577}

\bibitem[\protect\citeauthoryear{{Psaltis}}{{Psaltis}}{2008}]{Psaltis2008}
{Psaltis} D.,  2008, \mn@doi [Living Reviews in Relativity]
  {10.12942/lrr-2008-9}, 11, 1433

\bibitem[\protect\citeauthoryear{{Rangel}, {Nandra}, {Laird}  \&
  {Orange}}{{Rangel} et~al.}{2013}]{Rangel2013}
{Rangel} C.,  {Nandra} K.,  {Laird} E.~S.,   {Orange} P.,  2013, \mn@doi
  [\mnras] {10.1093/mnras/sts256}, \href
  {http://adsabs.harvard.edu/abs/2013MNRAS.428.3089R} {428, 3089}

\bibitem[\protect\citeauthoryear{{Rau} et~al.,}{{Rau} et~al.}{2013}]{Rau2013}
{Rau} A.,  et~al., 2013, arXiv e-prints, \href
  {https://ui.adsabs.harvard.edu/abs/2013arXiv1308.6785R} {p. arXiv:1308.6785}

\bibitem[\protect\citeauthoryear{{Rees}}{{Rees}}{1984}]{Rees1984}
{Rees} M.~J.,  1984, \mn@doi [\araa] {10.1146/annurev.aa.22.090184.002351},
  \href {https://ui.adsabs.harvard.edu/abs/1984ARA&A..22..471R} {22, 471}

\bibitem[\protect\citeauthoryear{{Reynolds} \& {Nowak}}{{Reynolds} \&
  {Nowak}}{2003}]{Reynolds2003}
{Reynolds} C.~S.,  {Nowak} M.~A.,  2003, \mn@doi [\physrep]
  {10.1016/S0370-1573(02)00584-7}, \href
  {http://adsabs.harvard.edu/abs/2003PhR...377..389R} {377, 389}

\bibitem[\protect\citeauthoryear{{Ricci} et~al.,}{{Ricci}
  et~al.}{2017}]{Ricci2017}
{Ricci} C.,  et~al., 2017, \mn@doi [\nat] {10.1038/nature23906}, \href
  {https://ui.adsabs.harvard.edu/abs/2017Natur.549..488R} {549, 488}

\bibitem[\protect\citeauthoryear{{Risaliti} \& {Elvis}}{{Risaliti} \&
  {Elvis}}{2004}]{Risaliti2004}
{Risaliti} G.,  {Elvis} M.,  2004, in {Barger} A.~J.,  ed.,  Astrophysics and
  Space Science Library Vol. 308, Supermassive Black Holes in the Distant
  Universe. p.~187 (\mn@eprint {} {astro-ph/0403618}),
  \mn@doi{10.1007/978-1-4020-2471-9_6}

\bibitem[\protect\citeauthoryear{{Risaliti} et~al.,}{{Risaliti}
  et~al.}{2013}]{Risaliti2013}
{Risaliti} G.,  et~al., 2013, \mn@doi [\nat] {10.1038/nature11938}, \href
  {https://ui.adsabs.harvard.edu/abs/2013Natur.494..449R} {494, 449}

\bibitem[\protect\citeauthoryear{Robert, Chopin  \& Rousseau}{Robert
  et~al.}{2009}]{Robert2009}
Robert C.~P.,  Chopin N.,   Rousseau J.,  2009, \mn@doi [Statist. Sci.]
  {10.1214/09-STS284}, 24, 141

\bibitem[\protect\citeauthoryear{{Robitaille} \& {Bressert}}{{Robitaille} \&
  {Bressert}}{2012}]{Robitaille2012}
{Robitaille} T.,  {Bressert} E.,  2012, {APLpy: Astronomical Plotting Library
  in Python}, Astrophysics Source Code Library (\mn@eprint {ascl} {1208.017})

\bibitem[\protect\citeauthoryear{{Ross} \& {Fabian}}{{Ross} \&
  {Fabian}}{1993}]{Ross1993}
{Ross} R.~R.,  {Fabian} A.~C.,  1993, \mn@doi [\mnras]
  {10.1093/mnras/261.1.74}, \href
  {https://ui.adsabs.harvard.edu/abs/1993MNRAS.261...74R} {261, 74}

\bibitem[\protect\citeauthoryear{{Salvato} et~al.,}{{Salvato}
  et~al.}{2009}]{Salvato2009}
{Salvato} M.,  et~al., 2009, \mn@doi [\apj] {10.1088/0004-637X/690/2/1250},
  \href {http://adsabs.harvard.edu/abs/2009ApJ...690.1250S} {690, 1250}

\bibitem[\protect\citeauthoryear{{Salvato} et~al.,}{{Salvato}
  et~al.}{2011}]{Salvato2011}
{Salvato} M.,  et~al., 2011, \mn@doi [\apj] {10.1088/0004-637X/742/2/61}, \href
  {http://adsabs.harvard.edu/abs/2011ApJ...742...61S} {742, 61}

\bibitem[\protect\citeauthoryear{{Shakura} \& {Sunyaev}}{{Shakura} \&
  {Sunyaev}}{1973}]{Shakura1973}
{Shakura} N.~I.,  {Sunyaev} R.~A.,  1973, \aap, \href
  {https://ui.adsabs.harvard.edu/abs/1973A&A....24..337S} {500, 33}

\bibitem[\protect\citeauthoryear{{Stan Development Team}}{{Stan Development
  Team}}{2014}]{stan-software:2014}
{Stan Development Team} 2014, Stan: A C++ Library for Probability and Sampling,
  Version 2.2, \url {http://mc-stan.org/}

\bibitem[\protect\citeauthoryear{{Steffen}, {Barger}, {Cowie}, {Mushotzky}  \&
  {Yang}}{{Steffen} et~al.}{2003}]{Steffen2003}
{Steffen} A.~T.,  {Barger} A.~J.,  {Cowie} L.~L.,  {Mushotzky} R.~F.,   {Yang}
  Y.,  2003, \mn@doi [\apjl] {10.1086/379142}, \href
  {https://ui.adsabs.harvard.edu/abs/2003ApJ...596L..23S} {596, L23}

\bibitem[\protect\citeauthoryear{{Streblyanska}, {Hasinger}, {Finoguenov},
  {Barcons}, {Mateos}  \& {Fabian}}{{Streblyanska}
  et~al.}{2005}]{Streblyanska2005}
{Streblyanska} A.,  {Hasinger} G.,  {Finoguenov} A.,  {Barcons} X.,  {Mateos}
  S.,   {Fabian} A.~C.,  2005, \mn@doi [\aap] {10.1051/0004-6361:20041977},
  \href {http://adsabs.harvard.edu/abs/2005A%26A...432..395S} {432, 395}

\bibitem[\protect\citeauthoryear{{Tanaka} et~al.,}{{Tanaka}
  et~al.}{1995}]{Tanaka1995}
{Tanaka} Y.,  et~al., 1995, \mn@doi [\nat] {10.1038/375659a0}, \href
  {http://adsabs.harvard.edu/abs/1995Natur.375..659T} {375, 659}

\bibitem[\protect\citeauthoryear{{Ueda}, {Akiyama}, {Ohta}  \& {Miyaji}}{{Ueda}
  et~al.}{2003}]{Ueda2003}
{Ueda} Y.,  {Akiyama} M.,  {Ohta} K.,   {Miyaji} T.,  2003, \mn@doi [\apj]
  {10.1086/378940}, \href
  {https://ui.adsabs.harvard.edu/abs/2003ApJ...598..886U} {598, 886}

\bibitem[\protect\citeauthoryear{{Vianello}}{{Vianello}}{2018}]{Vianello2018}
{Vianello} G.,  2018, \mn@doi [\apjs] {10.3847/1538-4365/aab780}, \href
  {https://ui.adsabs.harvard.edu/abs/2018ApJS..236...17V} {236, 17}

\bibitem[\protect\citeauthoryear{{Volonteri}, {Sikora}, {Lasota}  \&
  {Merloni}}{{Volonteri} et~al.}{2013}]{Volonteri2013}
{Volonteri} M.,  {Sikora} M.,  {Lasota} J.~P.,   {Merloni} A.,  2013, \mn@doi
  [\apj] {10.1088/0004-637X/775/2/94}, \href
  {https://ui.adsabs.harvard.edu/abs/2013ApJ...775...94V} {775, 94}

\bibitem[\protect\citeauthoryear{{Xue}, {Luo}, {Brandt}, {Alexander}, {Bauer},
  {Lehmer}  \& {Yang}}{{Xue} et~al.}{2016}]{Xue2016}
{Xue} Y.~Q.,  {Luo} B.,  {Brandt} W.~N.,  {Alexander} D.~M.,  {Bauer} F.~E.,
  {Lehmer} B.~D.,   {Yang} G.,  2016, VizieR Online Data Catalog, \href
  {http://adsabs.harvard.edu/abs/2016yCat..22240015X} {222}

\bibitem[\protect\citeauthoryear{{Yaqoob} \& {Padmanabhan}}{{Yaqoob} \&
  {Padmanabhan}}{2004}]{Yaqoob2004}
{Yaqoob} T.,  {Padmanabhan} U.,  2004, \mn@doi [\apj] {10.1086/381731}, \href
  {https://ui.adsabs.harvard.edu/abs/2004ApJ...604...63Y} {604, 63}

\bibitem[\protect\citeauthoryear{{de La Calle P{\'e}rez} et~al.,}{{de La Calle
  P{\'e}rez} et~al.}{2010}]{delaCalle2010}
{de La Calle P{\'e}rez} I.,  et~al., 2010, \mn@doi [\aap]
  {10.1051/0004-6361/200913798}, \href
  {https://ui.adsabs.harvard.edu/abs/2010A&A...524A..50D} {524, A50}

\makeatother
\end{thebibliography}



\appendix

\section{Hierarchical Bayesian model to infer the intrinsic R distribution}

\label{HBM}

\begin{figure}
\includegraphics[width=0.5\textwidth]{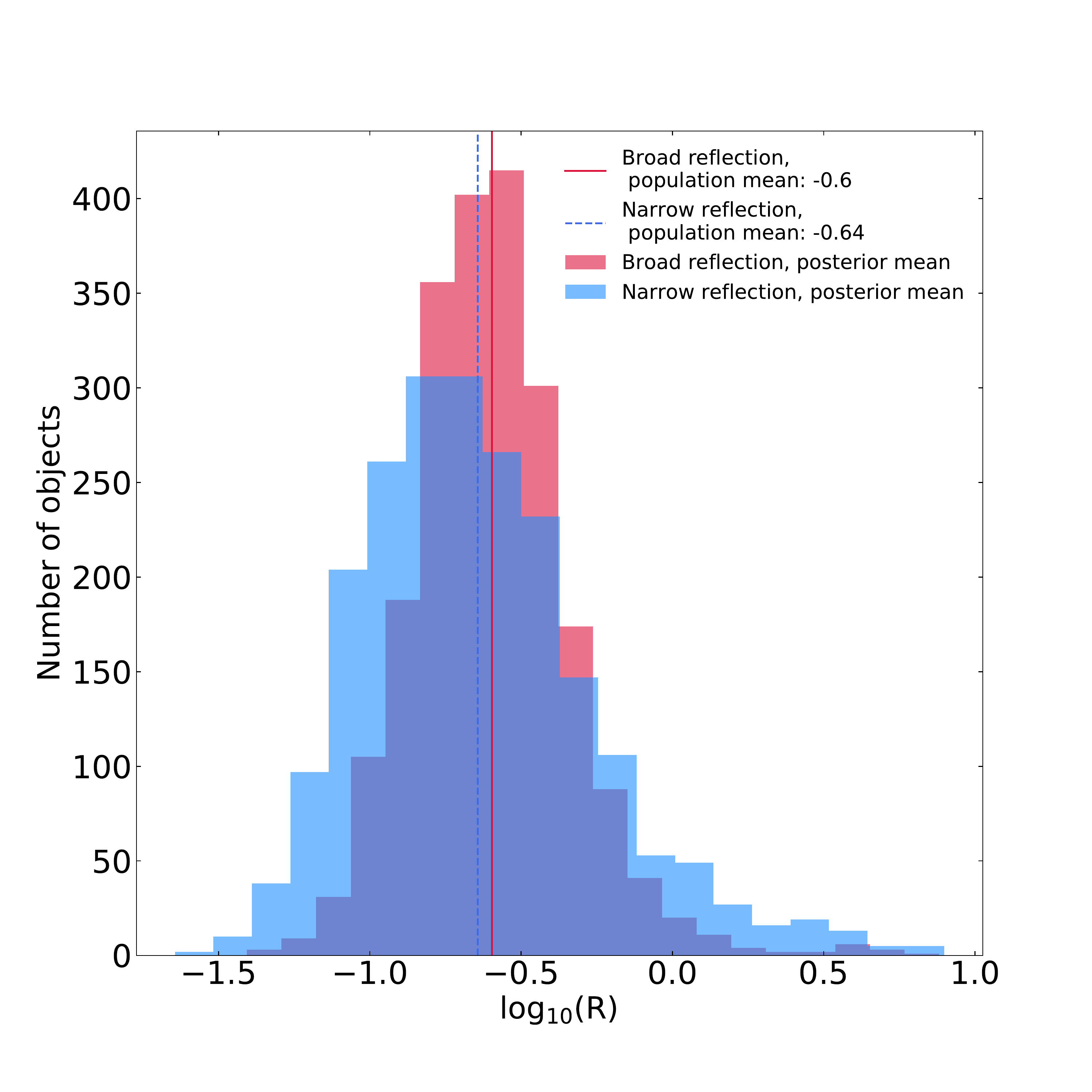}
\caption{Histograms of the mean of the posterior distributions of the R values for narrow and broad reflection components for every object. The vertical lines show the population mean (i.e. the mean of the posterior means) for narrow (red) and broad (blue dashed) reflection. }
\label{fig:Rmean}
\end{figure}

\begin{figure*}
\centering
\includegraphics[scale=0.3]{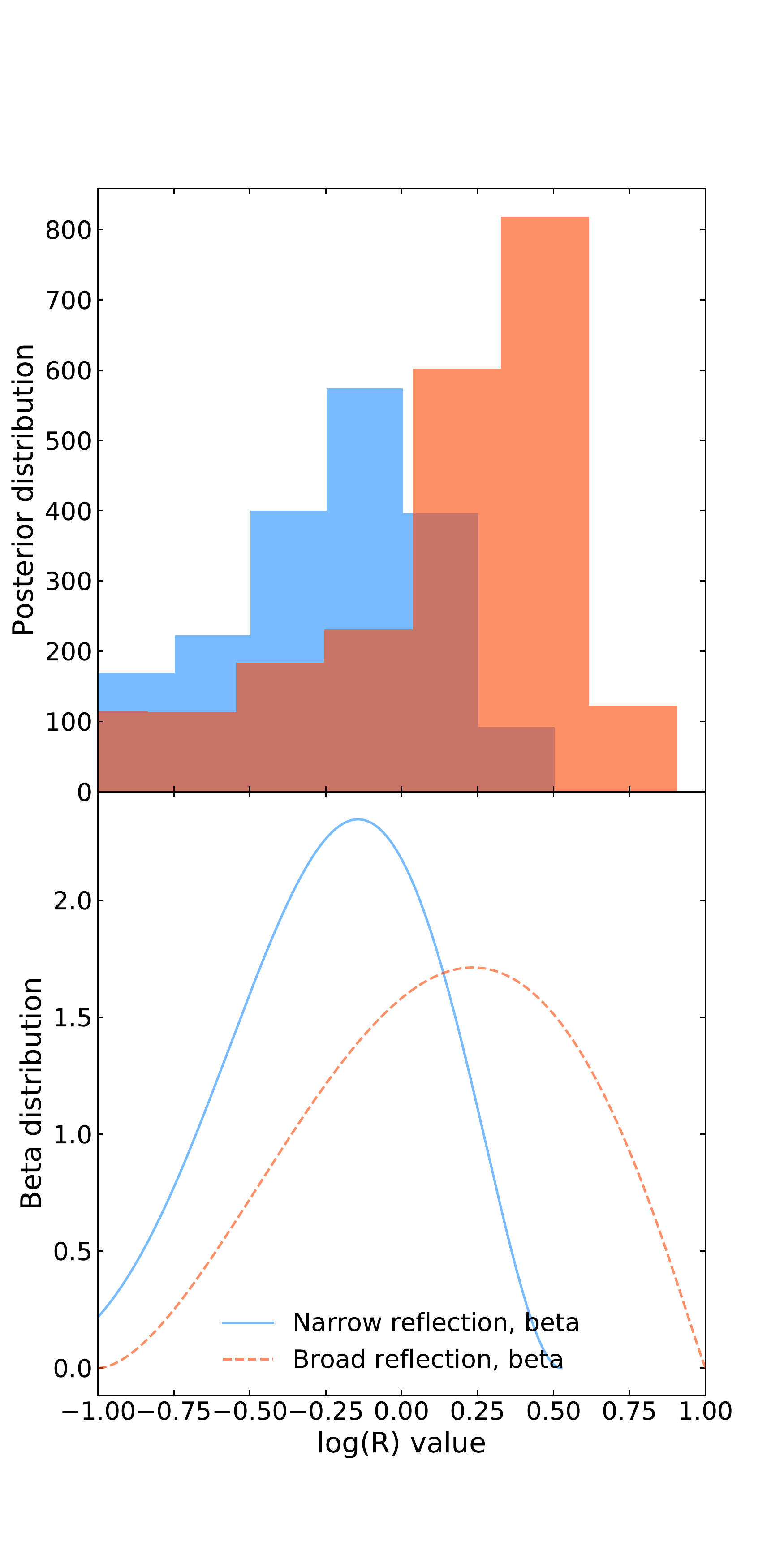}%
\includegraphics[scale=0.3]{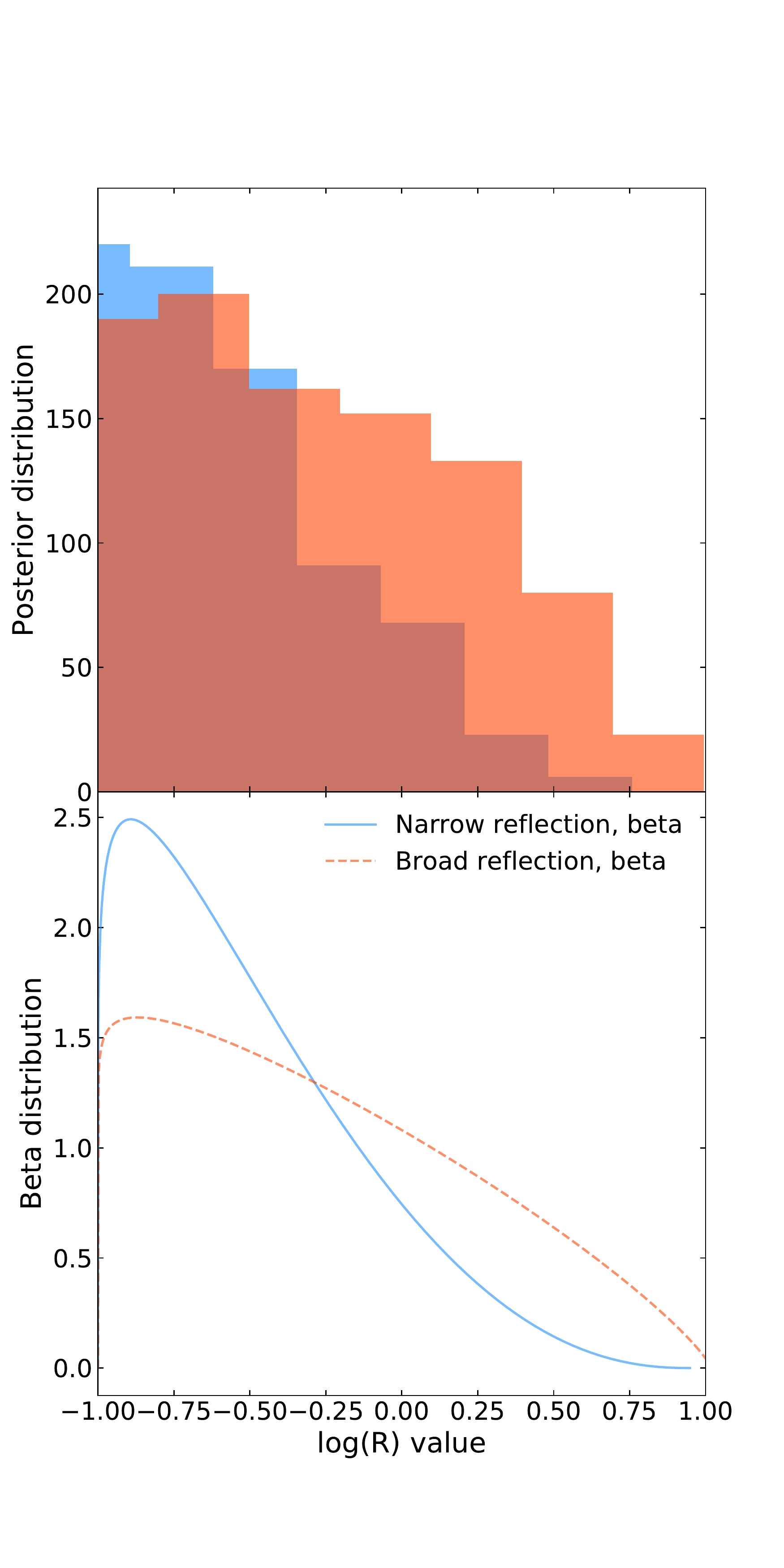}%
\includegraphics[scale=0.3]{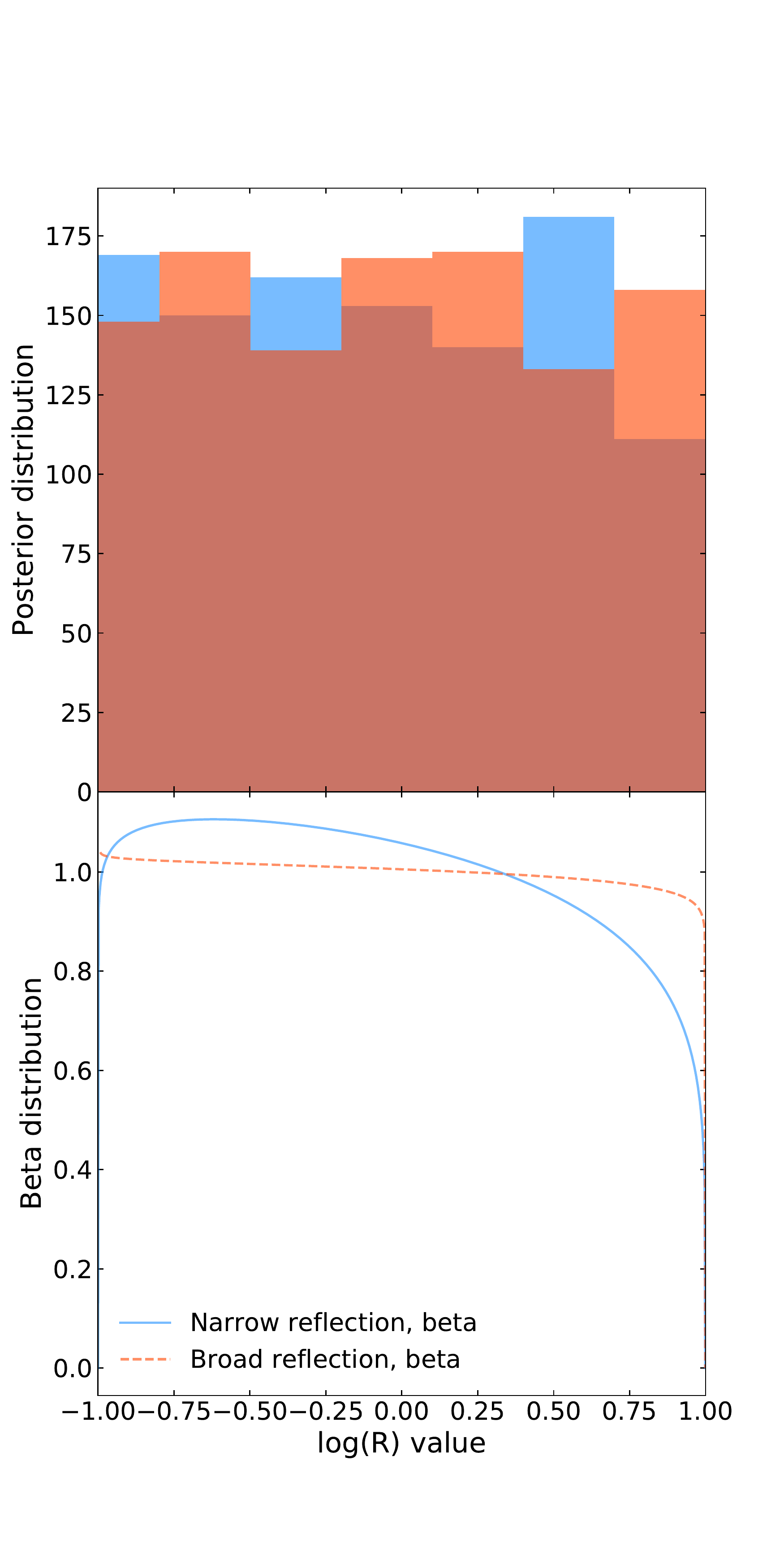}%
\caption{Three examples of the shapes of posterior distribution (\textit{top}) of the log R parameter and the fitted beta distribution (\textit{bottom}). The panels show a well constrained example (\textit{left}), an upper limit with the posterior concentrated in the lower half of the prior space (\textit{middle}) and a poorly constrained case (\textit{right}), where the posterior has approximately the shape of the prior distribution (uniform). The three objects whose posterior we show here are source 133, 225 and 345 from the CDFS 7Ms survey.}
\label{fig:beta}
\end{figure*}

In this work, we have derived posterior distributions for each model parameter over a large sample ($N>1000$ objects). The parameter of foremost interest in this work is R, which we derived from fitting a model to X-ray spectra of AGN using BXA. 

The posterior distributions are diverse and often wide, as illustrated in Figure \ref{fig:Rmean}. The question is now how to combine these uncertain posterior probability distributions to infer the intrinsic distribution.

Hierarchical Bayesian modelling (HBM) can use the information we have on the single object (the posterior distribution) to infer the features of the parent population from which the objects are drafted \citep{Betancourt2015}. In typical HBMs, the constraints per-object (e.g., its $R$ posterior) is solved simultaneously with the parameters of the population distribution (e.g., the mean and standard deviation R distribution of the population). This is a complex, very high-dimensional (>1000-dimensional) problem. In practice, we address this problem in two simplified ways. They are both based on the same probabilistic model laid out below, but make use of the $R$ posteriors already derived under flat priors. In one approach (Section~\ref{appsec:ultranest}), we use the posterior distributions as weights in numerical integration. In another approach, we solve the high-dimensional problem directly with Hamiltonian Monte Carlo (Section~\ref{appsec:stan}), but employ auxiliary distributions that approximate the posterior. The former approach has the benefit of not assuming a shape of the distributions. The latter approach avoids numerical sampling issues. In practice, the two methods of Section~\ref{appsec:stan} and \ref{appsec:ultranest} show consistent results, which gives confidence in the method. The presented tools for Hierarchical Bayesian modeling are thus powerful and robust for inferring the intrinsic distribution given a large number of uncertain measurements, including upper limits.


To derive the probabilistic model, we first assume that the parent distribution of the $\mathrm{log(R)}$ parameter is distributed as a Gaussian $\mathrm{ N(log R|\mu, \sigma)}$ with unknown mean $\mathrm{\mu}$ and standard deviation $\mathrm{\sigma}$. Adopting a different parent distribution shape, such as a skewed normal or a Beta distribution did not change the results significantly. For each object, we also have an a priori (before considering the data) unknown parameter R. We have already constrained this, as encoded in the posterior distributions $\mathrm{P(log R| D_{i})}$ for each object. We can thus write the combined likelihood for a single object as:

\begin{equation}
    \mathrm{\int P(log R| D_{i}) N(log R|\mu, \sigma) dlogR}.
    \label{eq:prob}
\end{equation}

Because the same parent distribution should hold for all objects, we multiply their probabilities and find the HBM likelihood:

\begin{equation}
    \mathrm{ \mathcal{L}= \prod_i \int P(log R| D_{i}) N(log R|\mu, \sigma) dlogR}.
    \label{eq:lik}
\end{equation}

Reusing the derived per-object posterior works here because we have adopted wide priors that are uniform over the integration variable in Eq.\ref{eq:lik} (log R).

After adopting priors on $\mu$ (uniform) and $\sigma$ (log-uniform), this forms a $2+N$-dimensional Bayesian inference problem. To derive posterior distributions on $\mu$ and $\sigma$, we use two techniques explained in the following sections.

\subsection{Hierarchical Bayesian Model inference with Stan}
\label{appsec:stan}

One way to solve Eq.\ref{eq:lik} is to fit for all $N+2$ parameters simultaneously. This requires advanced Hamiltonian Monte Carlo techniques which rely on likelihood gradients to navigate the search space.
One issue is that we do not want to refit the spectra in this process.
Therefore, to still allow each per-object $R$ to vary according to its spectral constraints, we adopt an analytic approximation to its posterior.

We first fit the posterior distributions of every object with a beta distribution. The free parameters are the shape parameters $\alpha$ and $\beta$ and the location and width of the distributions. Since the parameter range from $\mathrm{-2<log(R)<-1}$ does not have much physical sense and reflection fractions below 0.1 are virtually indistinguishable, we constrain the fit distributions to lie between -1 and 1.
The posterior shapes differ depending on whether the parameter is well constrained, not constrained or an upper/lower limit (see Figure \ref{fig:beta}, \textit{left} panels).
Thus, we choose the beta distribution because it is flexible enough to fit reliably distributions with different shapes. The bottom panels of Figure \ref{fig:beta} show our best-fit beta approximations. 

Next, we implement with the Hamiltonian Monte Carlo framework \textsc{Stan}\footnote{See \url{https://pystan.readthedocs.io/en/latest/}.} \citep{stan-software:2014} a model that reads the parameters of all Beta distributions (vectors of $\alpha$, $\beta$, location and scale). The model (Stan code in Listing \ref{code}) has free R parameters, which both follow these distributions and a parent normal distribution. The MCMC algorithm then simultaneously determines the posterior of the parameters (mean and sigma) of the normal parent distribution and that of the R values.

For the sample with S/N>7 we obtain a mean of $\mathrm{log(R_{pex}) = -0.53}$ with spread $\mathrm{\sigma_{log(R_{pex})} = 0.2}$ for the narrow reflection component and mean of $\mathrm{log(R_{blur}) = -0.57}$ with spread of $\mathrm{\sigma_{log(R_{blur})} = 0.14}$ for the blurred reflection component (see Figure \ref{fig:corner}). For comparison, if we average the means of every single posterior distribution of the R values we obtain a mean of $\mathrm{log(R_{pex}) = -0.64}$ with spread of $\mathrm{\sigma_{log(R_{pex})} = 0.38}$ for the narrow reflection component and mean of $\mathrm{log(R_{blur}) = -0.6}$ with spread of $\mathrm{\sigma_{log(R_{blur})} = 0.26}$ for the blurred reflection component (see Figure \ref{fig:Rmean}).

This method takes into account the large parameter uncertainties and upper limits. Since many of the posterior distributions for the R value have the shape of an upper limit (see Figure \ref{fig:beta}, \textit{middle}), the mean of the population we obtain with a HBM is much smaller than the mean we would obtain by simply averaging the mean of every posterior distribution (see Figure \ref{fig:Rmean}, for the simple mean and Figure \ref{fig:corner} for the mean and sigma obtained with a HBM method). 

The values presented in Figures \ref{fig:Baldwin} and \ref{fig:BaldwinR} were calculated by applying the HBM to the subsamples of objects in 6 bins of luminosity, column density and redshift.

\begin{lstlisting}[caption={Stan definition of a HBM where the input data has the shape of a beta distribution with parameters a, b, loc and scale and the model to be fit is a normal with parameters mu and sigma.\label{code}}]
data {
    int<lower=0> N;
    vector[N] a;
    vector[N] b;
    vector[N] loc;
    vector[N] scale;
}
parameters {
    real<lower=-1, upper=1> mu;
    real<lower=-2, upper=2> logsigma;
    vector<lower=0, upper=1>[N] u;
}
transformed parameters {
    vector[N] x;
    real<lower=0> sigma;
    x = u .* scale + loc;
    sigma = pow(10, logsigma);
}
model {
    u ~ beta(a, b);
    x ~ normal(mu, sigma);
}
\end{lstlisting}

\subsection{Numerical Hierarchical Bayesian Model inference}
\label{appsec:ultranest}

\begin{figure}
\centering
\includegraphics[width=\columnwidth]{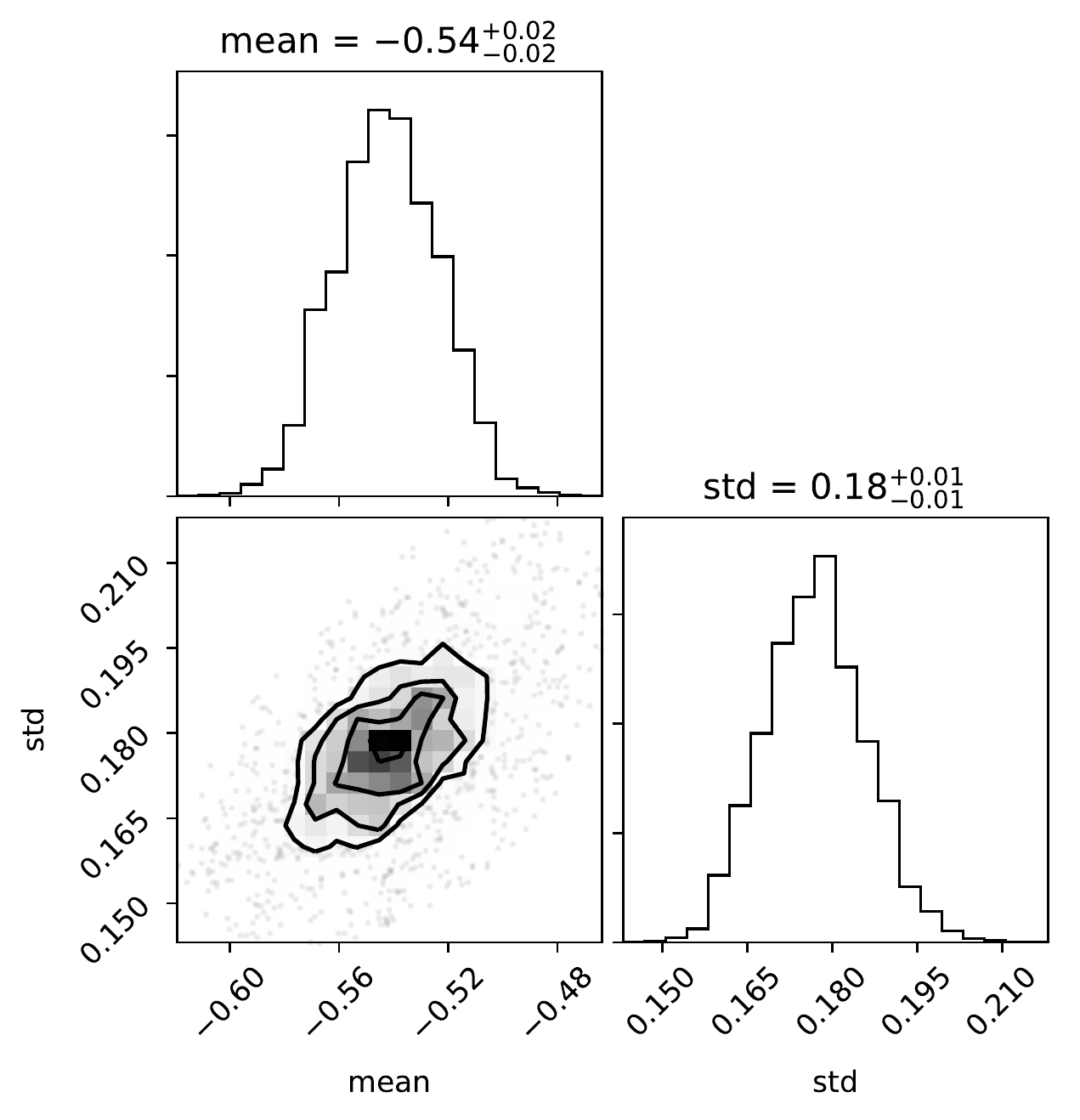}
\includegraphics[width=\columnwidth]{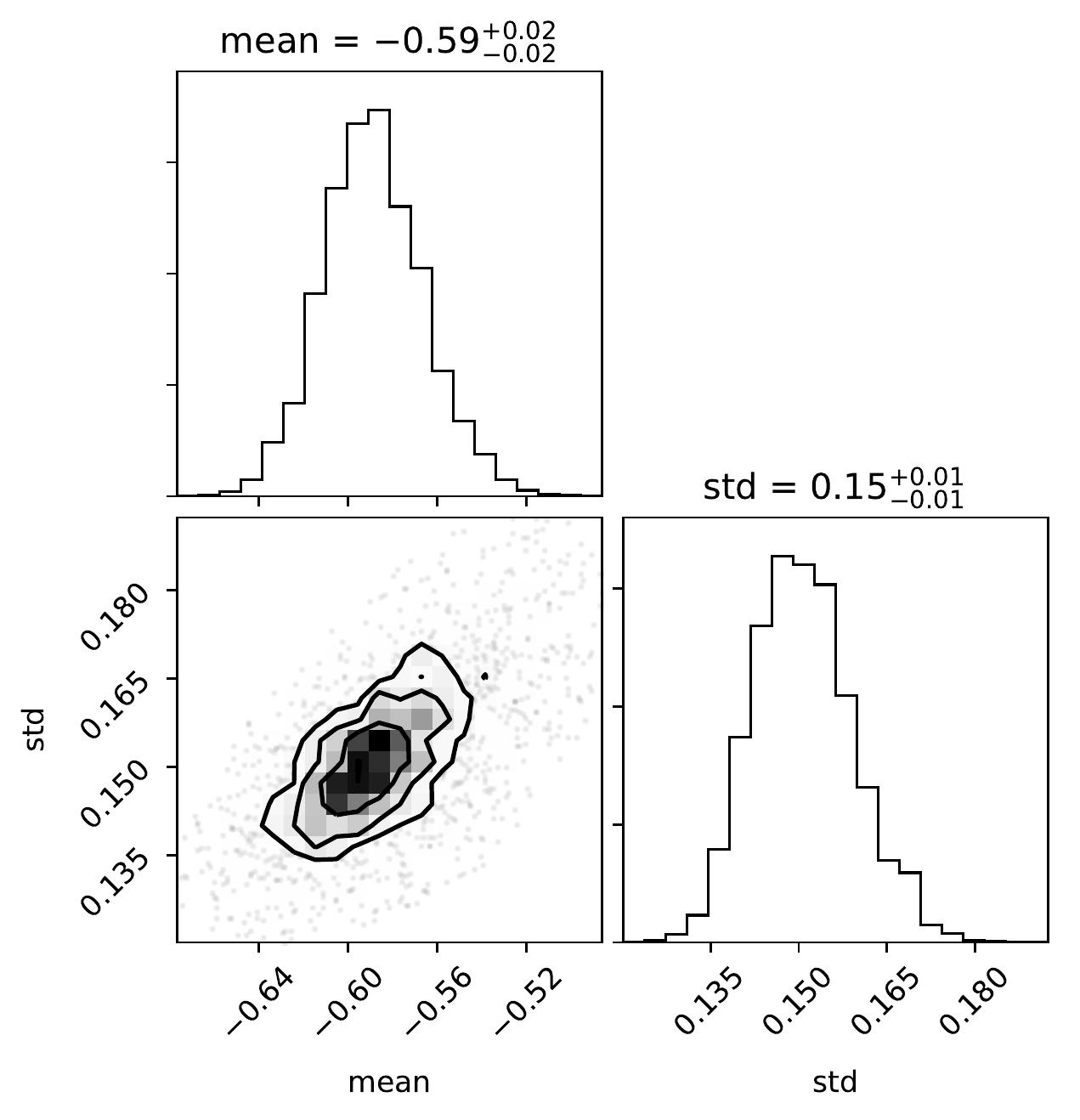}
\caption{Corner plots of the mean and sigma of the population of $\mathrm{log(R)}$ values for narrow (\textit{top}) and broad (\textit{bottom}) reflection component.}
\label{fig:ultranest}
\end{figure}

Another approach is to use importance sampling to numerically simplify the problem to 2 parameters. In practice, we already have posterior samples $\mathrm{R_{i,j}}$ for each object $i$ that approximate the (sometimes complex) posterior distributions. Therefore, we can write \ref{eq:lik}, dropping constant factors, with an importance sampling estimate:
\begin{equation}
    \mathrm{ \mathcal{L}(\mu, \sigma) \approx \prod_i \sum_j N(log R_{i,j}|\mu, \sigma)}.
    \label{eq:importance}
\end{equation}
When using too few posterior samples, this approach can induce numerical noise into the population posterior. Care has to be taken when this approach is used for multi-dimensional integrations \citep[see also][]{Buchner2015}.
Akin to cross-validation, this could be further improved by using sub-samples of the posterior samples in Eq.\ref{eq:importance}, and averaging the estimators. However, by varying the number of posterior samples used from hundreds to thousands, we verified that for our problem this Monte Carlo one-dimensional integration is stable.

The two-dimensional log-likelihood defined in Eq.\ref{eq:importance} is Monte Carlo sampled using \texttt{UltraNest}\footnote{See \url{https://johannesbuchner.github.io/UltraNest/index.html}}, a python nested sampling package developed in \citet{Buchner2019}. Figure \ref{fig:ultranest} shows the posterior of the mean and sigma of the normal distribution.
For the total sample with S/N>7 we obtain a mean of $\mathrm{log(R_{pex}) = -0.54}$ with spread $\mathrm{\sigma_{log(R_{pex})} = 0.18}$ for the narrow reflection component and mean of $\mathrm{log(R_{blur}) = -0.59}$ with spread of $\mathrm{\sigma_{log(R_{blur})} = 0.14}$ for the blurred reflection component.


\bsp	
\label{lastpage}
\end{document}